\pdfoutput=1

\documentclass[11pt,twoside,a4paper,cmspaper,final,collab]{cms-tdr}
\def\svnVersion{445219P}\def\cmsCernNoTag{CERN-EP-2017-269}\def\cmsCernDate{\today}\def\cmsMessage{Published in Physical Review D as \href{http://dx.doi.org/10.1103/PhysRevD.97.012007}{\doi{10.1103/PhysRevD.97.012007.}}}

\begin{document}\cmsNoteHeader{SUS-16-050}

\hyphenation{had-ron-i-za-tion}
\hyphenation{cal-or-i-me-ter}
\hyphenation{de-vices}
\RCS$Revision: 438348 $
\RCS$HeadURL: svn+ssh://svn.cern.ch/reps/tdr2/papers/SUS-16-050/trunk/SUS-16-050.tex $
\RCS$Id: SUS-16-050.tex 438348 2017-12-10 21:22:40Z alverson $
\newlength\cmsFigWidth
\ifthenelse{\boolean{cms@external}}{\setlength\cmsFigWidth{0.98\columnwidth}}{\setlength\cmsFigWidth{0.6\textwidth}}
\ifthenelse{\boolean{cms@external}}{\providecommand{\cmsLeft}{top\xspace}}{\providecommand{\cmsLeft}{left\xspace}}
\ifthenelse{\boolean{cms@external}}{\providecommand{\cmsRight}{bottom\xspace}}{\providecommand{\cmsRight}{right\xspace}}

\newcommand{\nbjets}{\ensuremath{N_{\PQb}}\xspace}
\newcommand{\njets}{\ensuremath{N_{\text{j}}}\xspace}
\newcommand{\ntops}{\ensuremath{N_{\PQt}}\xspace}
\newcommand{\MTTwo}{\ensuremath{m_{\text{T2}}}\xspace}
\newcommand{\mt}{\ensuremath{m_{\text{T}}}\xspace}
\newcommand{\htmiss}{\ensuremath{H_{\text{T}}^{\text{miss}}}\xspace}
\newcommand{\met}{\ptmiss}
\newcommand{\metv}{\ptvecmiss}
\newcommand{\znunu}{\ensuremath{\cPZ\to\PGn\PAGn}\xspace}
\newcommand{\znunujets}{\ensuremath{\cPZ(\PGn\PAGn)}\text{+jets}\xspace}
\newcommand{\Znunujets}{\znunujets}
\newcommand{\zll}{\ensuremath{\cPZ\to\ell\ell}}
\newcommand{\zjets}{{{\cPZ}\text{+jets}}\xspace}
\newcommand{\wjets}{{{\PW}\text{+jets}}\xspace}
\newcommand{\stopq}{\PSQt}
\newcommand{\gluino}{\PSg}
\newcommand{\topq}{\cPqt}
\newcommand{\lsp}{\PSGczDo}
\newcommand{\mlsp}{\ensuremath{m_{\lsp}}\xspace}
\newcommand{\qqprimebar}{\ensuremath{\PQq\PAQq^\prime}\xspace}
\newcommand{\ttbarW}{\ensuremath{\ttbar\PW}\xspace}
\newcommand{\ttbarZ}{\ensuremath{\ttbar\cPZ}\xspace}
\newcommand{\njetsisr}{\ensuremath{N_{\text{jet}}^{\text{ISR}}}\xspace}

\cmsNoteHeader{SUS-16-050}
\title{Search for supersymmetry in proton-proton collisions at \texorpdfstring{13\TeV}{13 TeV}
using identified top quarks}

\date{\today}

\abstract{
A search for supersymmetry is presented based on proton-proton collision events containing identified hadronically decaying top quarks, no leptons, and an imbalance \ptmiss in transverse momentum. The data were collected with the CMS detector at the CERN LHC at a center-of-mass energy of 13\TeV, and correspond to an integrated luminosity of 35.9\fbinv. Search regions are defined in terms of the multiplicity of bottom quark jet and top quark candidates, the \ptmiss, the scalar sum of jet transverse momenta, and the \MTTwo mass variable. No statistically significant excess of events is observed relative to the expectation from the standard model. Lower limits on the masses of supersymmetric particles are determined at 95\% confidence level in the context of simplified models with top quark production. For a model with direct top squark pair production followed by the decay of each top squark to a top quark and a neutralino, top squark masses up to 1020\GeV and neutralino masses up to 430\GeV are excluded. For a model with pair production of gluinos followed by the decay of each gluino to a top quark-antiquark pair and a neutralino, gluino masses up to 2040\GeV and neutralino masses up to 1150\GeV are excluded. These limits extend previous results.
}

\hypersetup{%
pdfauthor={CMS Collaboration},%
pdftitle={Search for supersymmetry in proton-proton collisions at 13 TeV
using identified top quarks},%
pdfsubject={CMS},%
pdfkeywords={CMS, physics, supersymmetry, top quark tagging}}

\maketitle

\section{Introduction}
\label{sec:Introduction}

The observation~\cite{Aad:2012tfa,Chatrchyan:2012ufa,Chatrchyan:2013lba}
of a Higgs boson (\PH) has been the most significant
discovery to date at the CERN LHC.
However, its relatively small mass of about 125\GeV~\cite{Aad:2015zhl}
can be understood in the context of the standard model (SM)
only through fine tuning of
the associated quantum loop corrections~\cite{Barbieri:1987fn}.
A compelling model that can account for the observed Higgs boson mass
without this fine tuning is the extension to the SM called supersymmetry
(SUSY)~\cite{Ramond,Golfand,Volkov,Wess:1974tw,Fayet,Barbieri:1982eh,Chamseddine,Hall,Kane}.
The main assertion of SUSY is the existence of one or more particles,
called superpartners,
for every SM particle,
where the spin of a superpartner differs from that of
its SM counterpart by a half integer.
The superpartners of quarks, gluons, and Higgs bosons are
squarks {\sQua}, gluinos {\PSg}, and higgsinos, respectively,
while neutralinos \PSGcz and charginos \PSGcpm
are mixtures of the superpartners
of electroweak and Higgs bosons.
In so-called natural models of SUSY~\cite{Papucci:2011wy},
the top squark,
bottom squark, gluino, and higgsinos are
required to have masses no larger,
and often much smaller,
than a few \TeV,
motivating searches for these particles
at the LHC.

In this paper we present a search for top squarks and gluinos.
The data were collected in 2016 by the CMS experiment at the LHC
and correspond to an integrated luminosity of 35.9\fbinv
of proton-proton ($\Pp\Pp$) collisions at
a center-of-mass energy of 13\TeV.
The search is performed in all-hadronic events with a
large imbalance \ptmiss in transverse momentum,
where by ``all-hadronic''
we mean that the final states are
composed solely of hadronic jets.
Recent searches for SUSY in a similar final state
are presented in
Refs.~\cite{Aaboud:2017ayj,Sirunyan:2017cwe,Sirunyan:2017kqq,Sirunyan:2017wif,Khachatryan:2017rhw}.
The current analysis is distinguished by the requirement that
identified (``tagged'') hadronically decaying top quarks be present.
It represents an extension,
using improved analysis techniques and a data sample 16 times larger,
of the study in Ref.~\cite{Khachatryan:2017rhw}.

In the search,
top squarks are assumed to be produced either
through the direct production of a top squark-antisquark pair
or in the decay of pair-produced gluinos.
They are assumed
to decay to the lightest neutralino $\lsp$---taken to be a
stable, weakly interacting, lightest SUSY particle (LSP)---and a quark.
Since the LSP interacts only weakly,
it does not produce a signal in the detector,
thus generating \ptmiss.
A novel top quark tagging algorithm is employed to identify hadronically
decaying top quarks produced in the decay chains.
The algorithm
makes use of the facts that a top quark essentially always decays
to a bottom quark and a {\PW} boson,
and that---in hadronic decays---the
{\PW} boson decays to a quark-antiquark (\qqprimebar) pair.
The algorithm recognizes three different types of
decay topology for the top quark.
In order of increasing Lorentz boost for the top quark, these are:
(i)~three distinct jets with no more than one of them identified
as a bottom quark jet (``$\cPqb$ jet''),
where two non-{\cPqb} jets arise from the
{\cPq} and {\cPaq$^\prime$} produced in the \PW~boson decay;
(ii)~two distinct jets, one of which corresponds to the {\cPqb} quark
and the other to the merged \qqprimebar decay products from the \PW~boson;
and (iii)~a single jet representing the merged decay products
of the {\cPqb} quark and {\PW} boson.
By accounting for these three different topologies,
the algorithm achieves high detection efficiency
over a wide range of top quark transverse momentum~\pt.

Events are selected that contain large \ptmiss,
at least four jets,
at least one identified $\cPqb$ jet,
at least one identified top quark,
and no identified leptons.
Search regions are defined based on the number \nbjets
of identified $\cPqb$ jets,
the number \ntops of top quark candidates,
the \ptmiss,
the scalar sum \HT of the \pt of jets,
and the \MTTwo~\cite{Lester:1999tx,Barr:2003rg} mass variable,
where \MTTwo is calculated using the reconstructed top quarks.

The largest source of SM background arises
from top quark-antiquark pair (\ttbar),
single top quark, and \wjets production,
namely from events in which a leptonically decaying
\PW~boson yields both a high-momentum neutrino,
generating \ptmiss,
and a charged lepton that is either not identified,
not reconstructed, or outside the analysis acceptance.
Another important source of background is \zjets production
followed by {\znunu} decay.
Quantum chromodynamics (QCD) multijet events,
namely events with multijet final states produced
exclusively through the strong interaction,
can contribute to the background if mismeasurement of jet \pt
yields large reconstructed \ptmiss or if a
semileptonically decaying charm or bottom hadron is produced.
Events with \ttbar production in which both top quarks decay
hadronically are indistinguishable from QCD multijet events
and are included in the QCD multijet background.
Because of the relatively small \ttbar cross section,
these \ttbar events constitute only a few percent of the
evaluated QCD multijet background.
Small sources of background include
multiple vector boson production and
events with a \ttbar pair
produced in association with a \cPZ~boson.

\section{Signal models}
\label{sec:models}

Signal scenarios for SUSY are considered in the context of simplified
models~\cite{Alwall:2008ag,Alwall:2008va,Alves:2011wf,Alves:2011sq,Chatrchyan:2013sza}.
For direct top squark pair production,
the simplified model denoted ``T2tt'' is examined.
In this model,
each top squark $\stopq$ decays to a top quark and the LSP:
$\stopq \to \topq \lsp$.
For top squark production through gluino decay,
the models described in the following two paragraphs are considered.

In the model denoted ``T1tttt,''
pair-produced gluinos each decay to an off-shell top squark
and an on-shell top quark.
The off-shell top squark decays to a top quark and the LSP.
The gluino decay is thus
$\gluino \to \cPqt \cPaqt \lsp$.
The T1tttt model provides sensitivity to situations
in which the top squark is too heavy to be produced directly
while the gluino is not.
In the ``T1ttbb'' model,
pair-produced gluinos each decay
via an off-shell top or bottom squark as
$\gluino \to\ttbar\PSGczDo$ (25\%),
$\gluino \to\cPaqt\cPqb\PSGcp_1$ or its charge conjugate (50\%),
or $\gluino \to\bbbar\PSGczDo$ (25\%),
where $\PSGcp_1$ is the lightest chargino.
The mass difference between the $\PSGcp_1$
and the LSP is taken to be $\Delta m(\PSGcp_1,\lsp)=5\GeV$.
Thus the $\PSGcp_1$ is taken to be nearly mass degenerate with the \lsp,
representing the expected situation should the
two particles appear within the same SU(2)
multiplet~\cite{Alves:2011wf}.
The $\PSGcp_1$ subsequently decays to the LSP and an off-shell $\PW$ boson.
The T1ttbb model provides sensitivity to mixed states of top and bottom squarks.

In the model denoted ``T5tttt,''
the mass difference between the top squark
and the LSP is $\Delta m(\stopq,\lsp) = 175\GeV$.
Pair-produced gluinos each decay to a top quark and an on-shell top squark.
The top squark decays to a top quark and the LSP.
This model provides sensitivity to a region that
is difficult to probe with the T2tt model because of
the similarity between the properties of T2tt signal
and \ttbar background events when $\Delta m(\stopq,\lsp)$
approximately equals the top quark mass~($m_\cPqt$).
The ``T5ttcc''  model is similar to the T5tttt model
except it assumes $\Delta m(\stopq,\lsp) = 20\GeV$
and the top squark decays to a charm quark and the LSP.
Note that decay to a charm quark and an LSP
represents the dominant decay mode of a top squark
when its decay to a top quark and an LSP is kinematically disallowed.
The choice of $\Delta m(\stopq,\lsp)$
has little effect on the final results for the T5ttcc model
(Section~\ref{sec:results})
so long as $\Delta m(\stopq,\lsp)$ remains below~$m_\cPqt$.
The T5ttcc model provides
sensitivity to scenarios in which the top squark is
kinematically unable to decay to an on-shell top quark.

\begin{figure*}[tbh]
  \centering
    \includegraphics[width=0.4\textwidth]{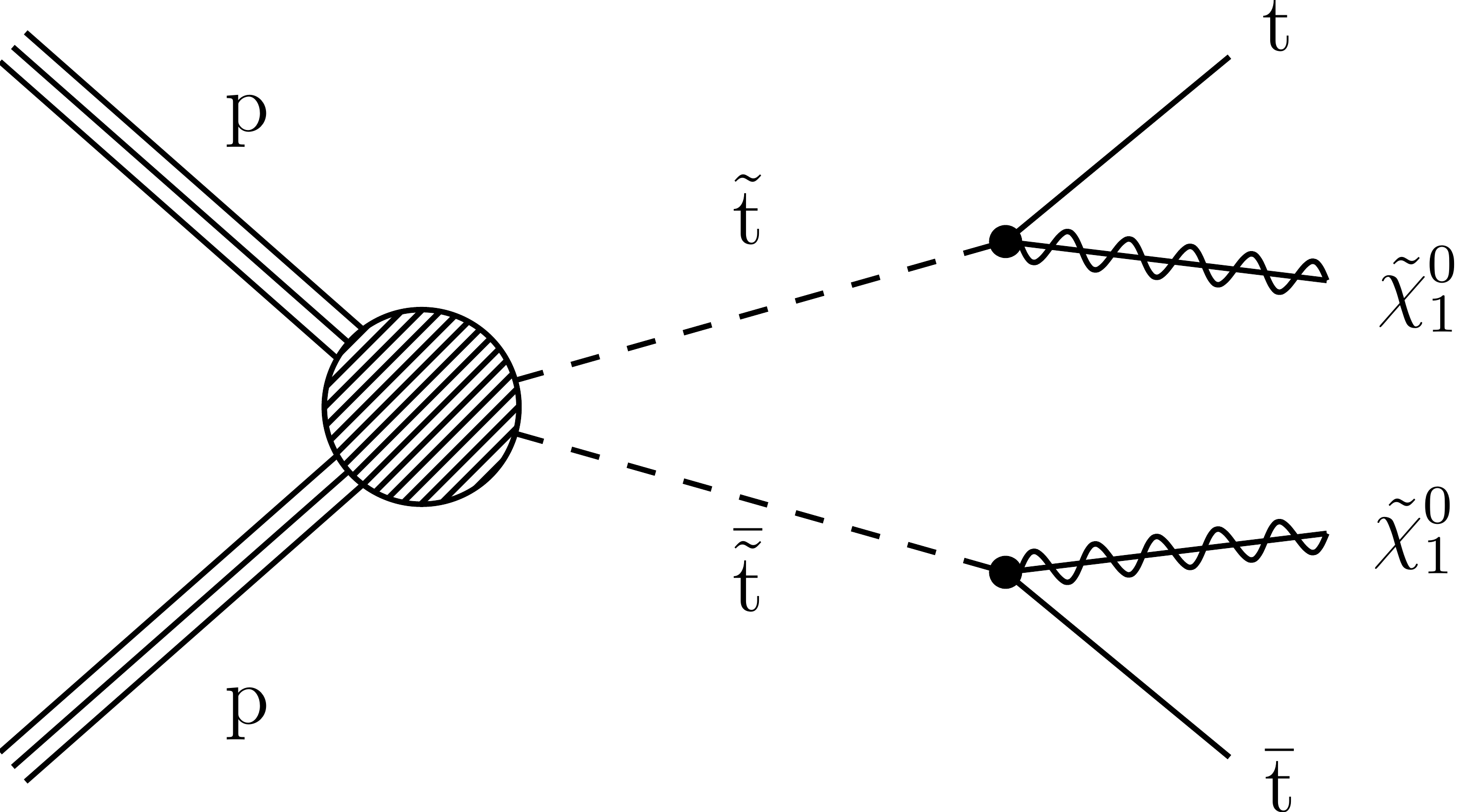}
    \includegraphics[width=0.4\textwidth]{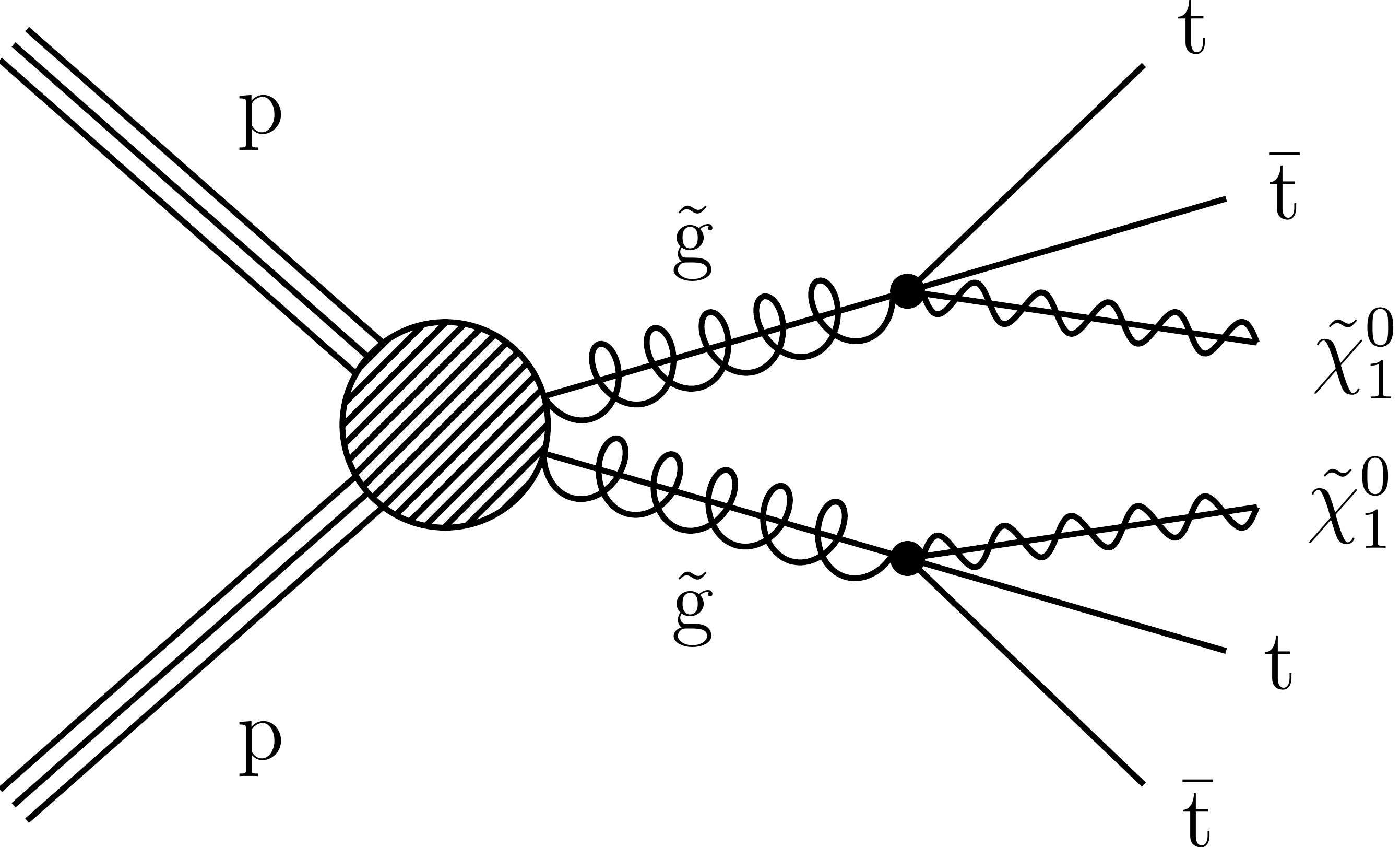} \\
    \includegraphics[width=0.4\textwidth]{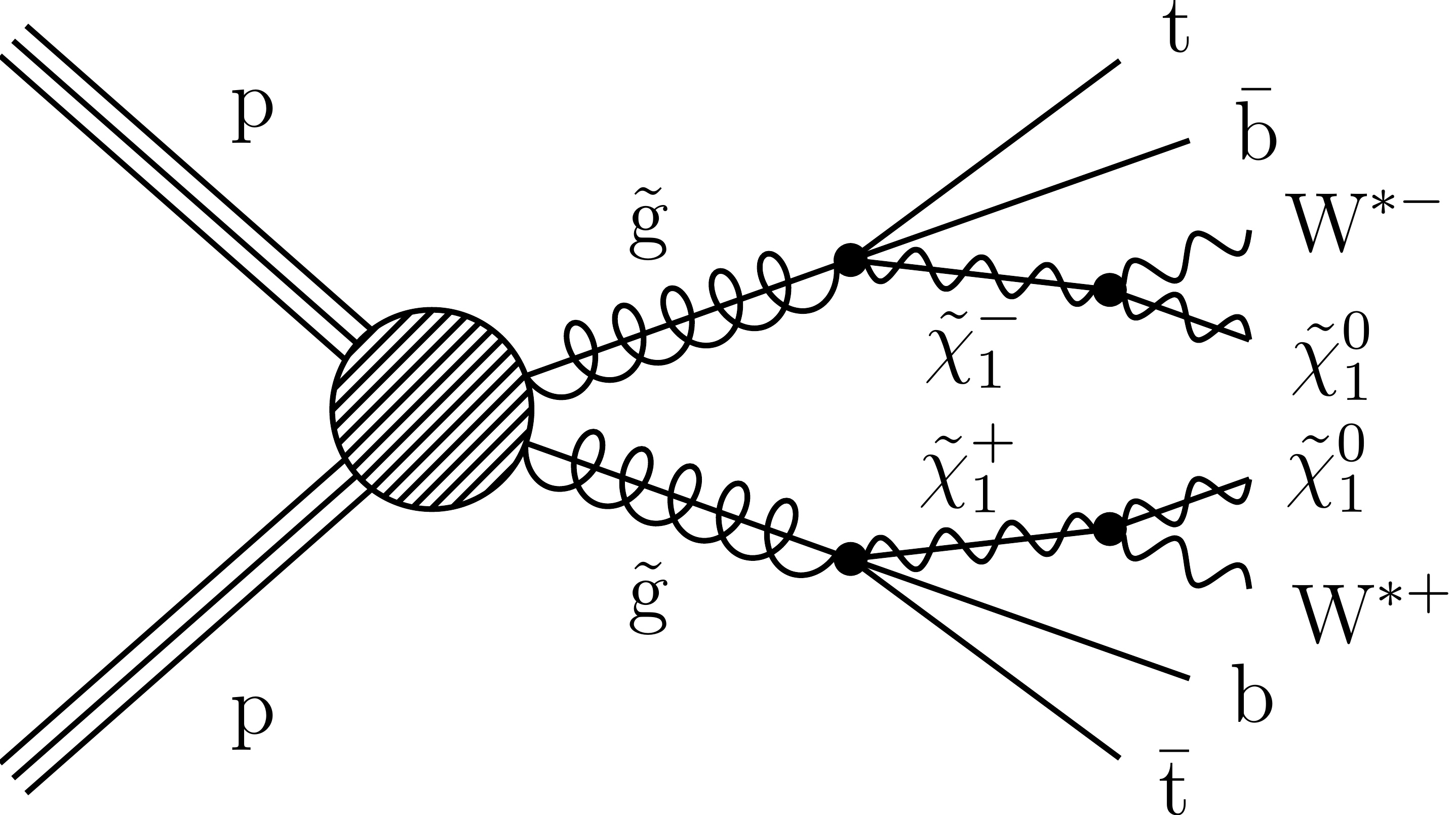}
    \includegraphics[width=0.4\textwidth]{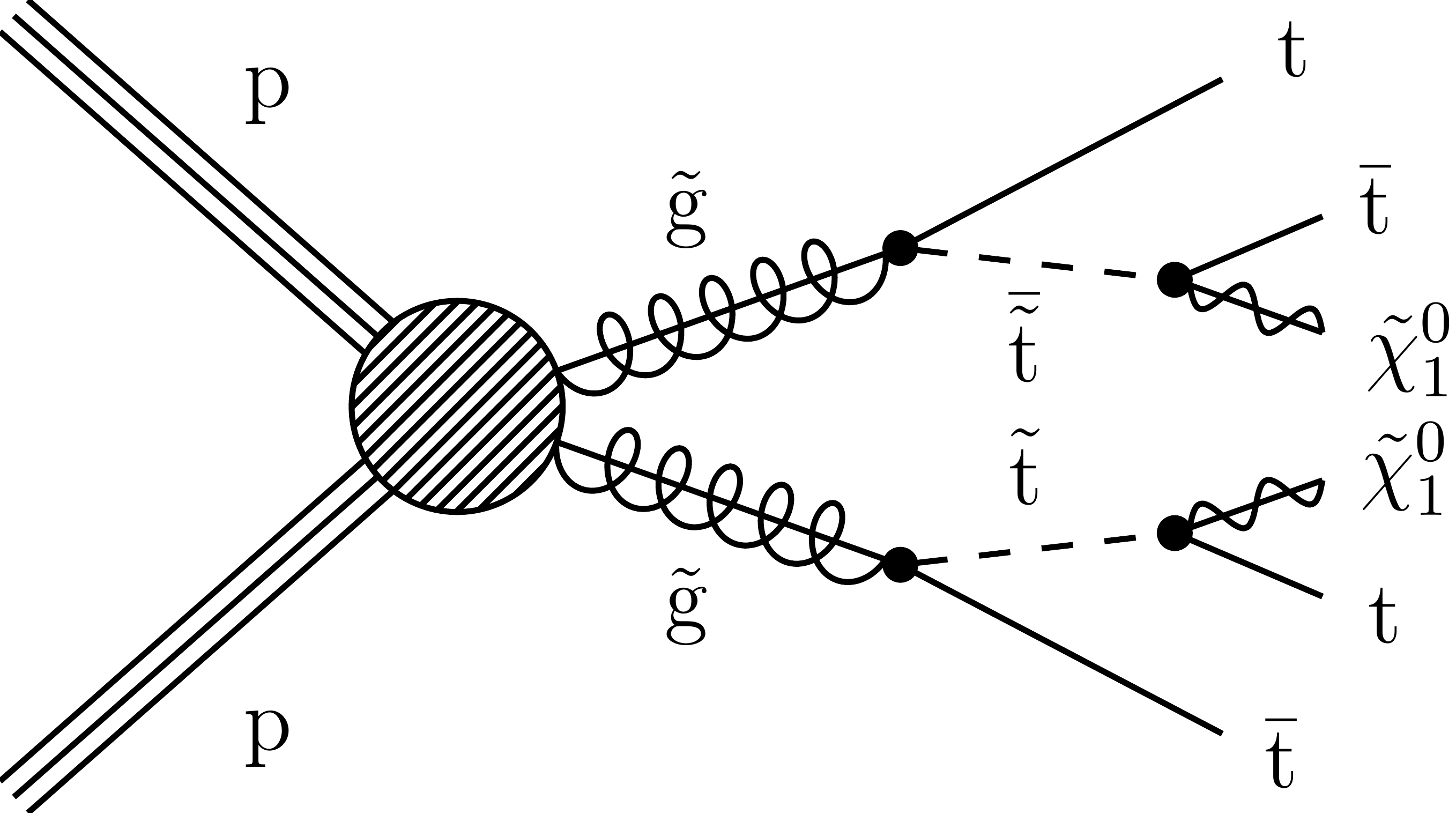} \\
    \includegraphics[width=0.4\textwidth]{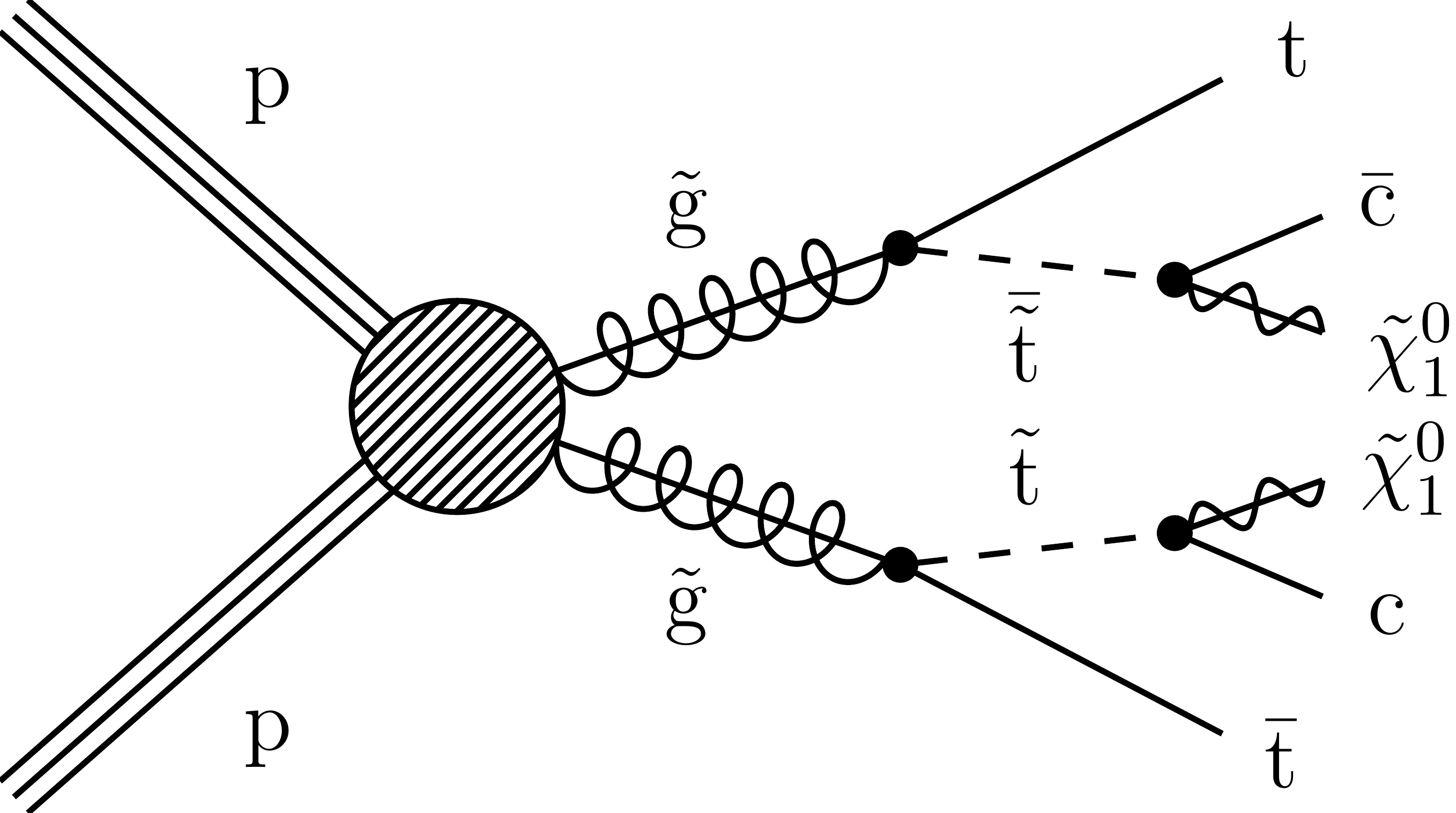}

   \caption{
   Diagrams representing the simplified models of direct
   and gluino-mediated top squark
   production considered in this study: the T2tt model (top left),
   the T1tttt model (top right), the T1ttbb model (middle left),
   the T5tttt (middle right), and the T5ttcc model (bottom).
}
  \label{fig:sms}
\end{figure*}

The signal scenarios are illustrated in Fig.~\ref{fig:sms}.
They exhibit common features,
such as the presence of multiple top quarks and two LSPs.

\section{The CMS detector}
\label{sec:CMSdetector}

The CMS detector is built around a superconducting
solenoid of 6\unit{m} internal diameter,
which provides a magnetic field of 3.8\unit{T}.
Within the solenoid volume are a silicon pixel and strip tracker,
a lead tungstate crystal electromagnetic calorimeter (ECAL),
and a brass and scintillator hadron calorimeter (HCAL).
The tracking detectors extend over the
pseudorapidity range $\abs{\eta}< 2.5$.
The ECAL and HCAL, each composed of a barrel and two endcap sections,
cover  $\abs{\eta}< 3.0$.
Forward calorimeters on each side of the interaction point
encompass $3.0<\abs{\eta}< 5.2$.
Muons are detected within $\abs{\eta}< 2.4$
by gas-ionization chambers embedded in a steel magnetic flux-return
yoke outside the solenoid.
A more detailed description of the CMS detector,
together with a definition of the
coordinate system used and the relevant kinematic variables,
can be found in Ref.~\cite{Chatrchyan:2008zzk}.

Events are selected using a two-level trigger system~\cite{Khachatryan:2016bia}.
The first level,
composed of custom hardware processors,
uses information from the calorimeters and muon detectors
to select events of interest at a rate of around 100\unit{kHz}.
The second level,
composed of a high-level processor farm,
decreases the event rate
to around 1\unit{kHz} before data storage.

For the present analysis,
events in the search regions are collected with a trigger
that requires $\ptmiss>100\GeV$ and $\htmiss>100\GeV$,
where \htmiss is the magnitude of the vector \pt sum of
jets reconstructed at the trigger level.
This trigger is fully efficient after application of
the event selection criteria described below.

\section{Event reconstruction}
\label{sec:eventreconstruction}

Events are reconstructed using the particle-flow (PF)
algorithm~\cite{Sirunyan:2017ulk},
which reconstructs charged hadrons, neutral hadrons, photons,
electrons, and muons using information from all subdetectors.
Electron and muon candidates are subjected to additional
requirements~\cite{Khachatryan:2015hwa,Chatrchyan:2013sba}
to improve their purity,
and are further required to have $\pt>10\GeV$
and to originate from within 2\unit{mm}
of the beam axis in the transverse plane.
Electron (muon) candidates must appear within $\abs{\eta}<2.5$ (2.4).
The missing transverse momentum \ptvecmiss in an event is given by
the negative of the vector \pt sum of all reconstructed objects.
Its magnitude is denoted \ptmiss.

All photons and neutral hadrons in an event,
together with charged particles that originate from the
primary interaction vertex,
are clustered into jets using the anti-\kt algorithm
with a distance parameter of 0.4 (AK4)~\cite{Cacciari:2008gp}.
The jets must satisfy a set of jet identification criteria as
specified in Ref.~\cite{CMS-PAS-JME-10-003}.
Neutral particles from overlapping $\Pp\Pp$ interactions (``pileup")
are subtracted on an
event-by-event basis using the \FASTJET technique~\cite{PU_JET_AREAS,JET_AREAS}.
Jets are corrected using factors from simulation to account
for detector response as a function of jet \pt and $\eta$.
Additional corrections account for residual
differences between simulation and data for
the jet energy and momentum scales~\cite{Khachatryan:2016kdb}.
Only jets with $\pt>30\GeV$ and either $\abs{\eta} < 2.4$ (tight)
or $\abs{\eta} < 5.0$ (loose) are retained.
The number of jets \njets in an event is defined to be the number
of tight AK4 jets.
The \HT variable is given by the scalar
sum of jet \pt over this same jet sample.

Bottom quark jets are identified ($\cPqb$ tagged)
by applying the combined secondary vertex
algorithm (CSVv2)~\cite{CMS-PAS-BTV-15-001,Chatrchyan:2012jua}
at the medium working point to tight AK4 jets.
The {\cPqb} quark identification efficiency ranges from 60 to 70\%
for jet \pt between 20 and 400\GeV.
The probability for a jet originating from a gluon or light-flavored quark
to be {\cPqb} tagged,
averaged over the jets in a sample
of \ttbar events, is 1.4\%~\cite{CMS-PAS-BTV-15-001}.

In addition to AK4 jets,
we define AK8 jets,
constructed by clustering PF objects using the anti-\kt algorithm
with a distance parameter of~0.8.
The AK8 jets are used in the top quark reconstruction procedure,
described in Section~\ref{sec:toptagger}.
Pileup contributions to AK8 jets are accounted for using the
``pileup per particle identification''~\cite{Bertolini:2014bba,CMS-PAS-JME-14-001} method,
by which each charged and neutral particle is weighted by a factor
representing its probability to originate from the primary interaction vertex
before the clustering is performed.
The AK8 jets are required to satisfy $\pt>200\GeV$.

\section{Lepton and track vetoes}
\label{sec:vetoes}

To obtain an all-hadronic event sample,
events with isolated electrons or muons are vetoed.
The isolation of electron and muon candidates is defined
as the scalar \pt sum of PF candidates in a cone of
radius $\Delta R = \sqrt{\smash[b]{(\Delta\eta)^2 + (\Delta\phi)^2}}$
around the candidate's trajectory,
where $\phi$ is the azimuthal angle and the sum excludes
the electron or muon candidate.
The cone size is 0.2 for $\pt\leq 50\GeV$,
0.05 for $\pt\geq 200\gev$,
and decreases in inverse proportion to the lepton \pt
for $50<\pt<200\GeV$.
This decrease in cone size with increasing lepton \pt
accounts for the greater collimation of a heavy object's decay products
as its Lorentz boost increases.
The isolation sum is corrected for contributions from pileup
using an estimate of the pileup energy in the cone~\cite{PU_JET_AREAS}.
Electron and muon candidates are considered to be isolated if their
relative isolation, \ie,
the ratio of the isolation sum to the candidate \pt,
is less than 0.1 and 0.2, respectively.

Events that survive the lepton veto
are subjected to an isolated charged-particle track veto.
This veto suppresses events with a hadronically decaying $\tau$ lepton
or with an isolated electron or muon not identified as such.
Tracks considered for this veto
must have $\pt > 5\GeV$, $\abs{\eta} < 2.5$,
and relative track isolation less than 0.2.
The relative track isolation is defined
analogously to the relative isolation of electrons and muons
but is computed using charged PF candidates only,
that appear within a fixed cone of $\Delta R = 0.3$ around the track.
To preserve signal efficiency,
the isolated-track veto is applied only if the transverse mass \mt~\cite{Arnison:1983rp}
of the isolated track-$\ptvecmiss$ system
is consistent with {\PW} boson decay,
namely $\mt<100\GeV$.
The isolated-track veto reduces background from events with
a leptonically decaying \PW~boson by about 40\%.

Following application of the above two vetoes,
a significant fraction of the remaining SM background arises from
events with a hadronically decaying $\tau$ lepton (\tauh).
A charged-hadron veto is applied to reduce this background.
The charged-hadron veto eliminates events that contain
an isolated PF charged hadron with $\pt>10\GeV$,
$\abs{\eta}<2.5$, and $\mt<100\GeV$.
To be considered isolated,
the relative isolation of the charged hadron,
defined as in the previous paragraph,
must be less than~0.1.

\section{Event simulation}
\label{sec:eventsimulation}

Samples of Monte Carlo (MC) simulated events are used to study the
properties of signal and background processes.
The {\MADGRAPH}5{\textunderscore}a{\MCATNLO}
2.2.2~\cite{Alwall:2014hca,Alwall:2007fs}
event generator at leading-order (LO) is used to describe signal events
and the SM production of $\ttbar$,
\wjets (with $\PW \to \ell\PGn$),
\zjets (with \znunu),
Drell--Yan (DY)+jets,
and QCD multijet events.
The \ttbar events are generated with up to three additional partons present
beyond those that participate in the hard scattering,
the signal events with up to two,
and the other processes with up to four.
The DY+jets events,
specifically events with the decay of a real or virtual {\cPZ} boson to a \MM pair,
are used as part of the procedure to evaluate background (Section~\ref{sec:Zinv}).
The generation of these processes is based on LO parton distribution
functions (PDFs) from NNPDF3.0~\cite{Ball:2014uwa}.
Single top quark events in the $\cPqt\PW$~channel
are generated with the next-to-leading order
(NLO) \POWHEG v2.0~\cite{Nason:2004rx,Frixione:2007vw,Alioli:2010xd,Re:2010bp}
program.
The following rare SM processes are considered:
$\ttbarZ$, $\ttbarW$, triboson, and ${\ttbar}\PH$ production,
generated at NLO with the
{\MADGRAPH}5{\textunderscore}a{\MCATNLO}
2.2.2~\cite{Alwall:2014hca,Frederix:2012ps} program using NLO NNPDF3.0 PDFs;
$\PW\cPZ$ and $\cPZ\cPZ$ production,
generated either with this same program or with the
\POWHEG program mentioned in the previous sentence
depending on the decay mode;
and $\PW\PW$ production,
generated with the \POWHEG program mentioned in the previous sentence.
Parton showering and hadronization are simulated for all MC samples
with the {\PYTHIA} v8.205~\cite{pythia8} program,
which uses the underlying event tune CUETP8M1~\cite{Khachatryan:2015pea}.

For simulated background processes,
the CMS detector response is based on the \GEANTfour package~\cite{Agostinelli:2002hh}.
Because of the intense computational requirements,
the detector response for simulated signal events is performed with
a fast simulation~\cite{fastsim},
which is tuned to provide results that are consistent with those
from the \GEANTfour-based simulation.
For all MC samples,
event reconstruction is performed in the same manner as for the data.

The signal production cross sections are calculated using
NLO plus next-to-leading logarithm (NLL) calculations~\cite{Borschensky:2014cia}.
The most precise cross section calculations currently available
are used to normalize the SM simulated samples,
corresponding to NLO or next-to-NLO accuracy in most
cases~\cite{Alwall:2014hca,Czakon:2011xx,Kant:2014oha,Aliev:2010zk,Gehrmann:2014fva,Campbell:1999ah,Campbell:2011bn,Li:2012wna}.

The simulated events are corrected for differences between simulation and data
in the {\cPqb} tagging efficiency,
the top quark tagging (Section~\ref{sec:toptagger}) efficiency,
and the electron and muon
identification and isolation selection efficiencies.
The corrections for the {\cPqb} tagging efficiency
are derived from multijet- and \ttbar-enriched event samples and are
parameterized in terms of the jet kinematics~\cite{CMS-PAS-BTV-15-001}.
The corrections for the top quark tagging efficiency are derived
from a single-muon \ttbar-enriched control sample and are
applied as a function of top quark~\pt.
The corrections for the electron and muon identification and isolation efficiencies
are determined from {\zll} events.

Simulated \ttbar and signal events are corrected
with scale factors to account for imperfect modeling of
initial-state radiation (ISR).
The ISR corrections are derived from a \ttbar-enriched
control sample containing two leptons
($\Pe\Pe$, $\Pgm\Pgm$, or $\Pe\Pgm$)
and two tagged {\cPqb} jets,
and are applied as a function of \njetsisr
up to $\njetsisr=6$,
where \njetsisr is the number of jets in the event
other than the two that are {\cPqb} tagged.
The correction is validated by applying it to simulation
in a \ttbar-dominated single-lepton control sample covering
various regions of phase space,
including regions with a large number of jets.
Agreement with data on the level of 20\% of the correction
is found in this control sample for key observables
such as the distribution in the number of jets.
To account for possible differences between \ttbar and signal events,
a conservative uncertainty of 50\% of the correction
is assigned to the scale factors,
both as applied to \ttbar and signal processes.

\section{Top quark reconstruction}
\label{sec:toptagger}

The top quark tagging algorithm is the central feature of our analysis.
It is designed to provide high reconstruction efficiency over the
full range of top quark \pt in the considered signal models.
A common strategy~\cite{CMS-PAS-JME-15-002,Aad:2016pux}
for tagging hadronically decaying top quarks is to
cluster jets with the AK8 algorithm
and then to test whether the jet is consistent with having
three subjets,
as expected for the
$\cPqt\to\cPqb\qqprimebar$ decay
of a highly Lorentz-boosted top quark.
Although these algorithms are efficient at large top quark \pt,
for $\pt<400\GeV$
top quarks are more efficiently reconstructed by combining
three individual AK4 jets,
an approach known as ``resolved'' top quark tagging.
To obtain high reconstruction efficiency over a wide range of top quark \pt,
we employ both types of algorithms and,
in addition, consider top quark decays in which
the decay products of the \PW~boson are contained within an AK8 jet.
To fully reconstruct the top quark in the latter case,
an AK8 jet corresponding to the \PW~boson decay is combined with an AK4 jet.

To identify high-\pt top quarks,
AK8 jets with $\pt>400\GeV$ are selected.
The mass of the jet is corrected with the soft-drop
method~\cite{Dasgupta:2013ihk, Larkoski:2014wba}
using angular exponent $\beta = 0$,
soft cutoff threshold $z_{\text{cut}}<0.1$,
and characteristic radius $R_{0} = 0.8$,
where the values of $\beta$, $z_{\text{cut}}$, and $R_{0}$ are
those recommended in Ref.~\cite{CMS-PAS-JME-16-003} for AK8 jets.
The soft-drop algorithm reclusters the AK8 jet into subjets
using the Cambridge--Aachen algorithm~\cite{Dokshitzer:1997in,Wobisch:1998wt}.
This reclustering removes soft radiation,
which can bias the jet mass determination.
To be considered as a top quark candidate,
the soft-drop mass must lie between 105 and 210\GeV.
The $N$-subjettiness variables $\tau_{N}$~\cite{Thaler:2010tr}
are used to determine the consistency of the jet with having three subjets.
More details on this algorithm can be found in Ref.~\cite{CMS-PAS-JME-15-002}.
To be consistent with having three subjets,
the requirement $\tau_3 / \tau_2 < 0.65$ is imposed.
This requirement is made on the basis of optimization
studies~\cite{CMS-PAS-JME-16-003}.

To avoid overlap between the top-tagged AK8 jets (denoted ``monojets'')
and the AK4 jets that are used to reconstruct resolved (``trijets'')
or partially merged (``dijets'') top quarks,
AK4 jets matched to the top-tagged AK8 jet are removed
from the list of AK4 jets used in the reconstruction
of the dijet and trijet categories.
An AK4 jet is considered matched if it lies within $\Delta R < 0.4$
of one of the soft-drop subjets of the tagged AK8 jet.

For the dijet category of top quark decays,
we employ a similar technique to identify the jet from
the hadronic {\PW} boson decay.
An AK8 jet with $\pt>200\GeV$ must have a soft-drop
corrected mass between 65 and 100\GeV.
To be consistent with having two subjets,
the requirement $\tau_2 / \tau_1 < 0.6$ is imposed.
This requirement corresponds to the ``high-purity pruning'' criterion
of Ref.~\cite{CMS-PAS-JME-16-003}.
The AK8 jet is combined with a loose AK4 jet
to form a top quark candidate.
The candidate must have a mass between 100 and 250\GeV,
both jets must appear within a cone of radius $\Delta R = 1$
around the direction of their summed \pt vector,
and the ratio of the soft-drop corrected AK8 jet mass
to the top quark candidate mass must lie between
$0.85 \,(m_\PW / m_\cPqt)$ and $1.25\, (m_\PW / m_\cPqt)$,
with $m_\PW$ the {\PW} boson mass.
If more than one top quark candidate is found using the same AK8 jet,
the combination with mass closest to $m_\cPqt$ is chosen.
The AK4 jet used to form the top quark candidate,
and all AK4 jets matched to within $\Delta R<0.4$ of the soft-drop subjets
from the AK8 jet,
are removed from the list used to reconstruct the trijet category.

The trijet sample of top quark candidates is formed by combining
three loose AK4 jets.
The three jets must appear within a cone of radius $\Delta R=1.5$
around the direction of their summed \pt vector,
no more than one of the three jets can be $\cPqb$ tagged,
and the trijet mass must lie between 100 and 250\GeV.
The cone size is chosen to be $\Delta R=1.5$
because the background becomes very large for larger $\Delta R$ values.
The final trijet top quark sample is defined by applying
the results of a random forest boosted decision tree~\cite{randomForests}
to the selected combinations.
The random forest is trained with simulation using
trijet combinations that satisfy the above criteria.
Simulated samples of \ttbar and \Znunujets events
are used for this purpose.
In the \ttbar simulation,
one top quark decays hadronically and the other semileptonically.
Signal top quarks are defined
as trijet combinations in the \ttbar simulation
for which each of the three jets is matched to a distinct
generator-level hadronically decaying top quark decay product
within $\Delta R< 0.4$,
and whose overall momentum is matched to the
generator-level top quark momentum within $\Delta R < 0.6$.
Background combinations are defined
as trijet combinations in the \ttbar sample
with no jet matched to a generator-level hadronically
decaying top quark decay product,
and as trijet combinations in the \Znunujets sample.
If more than one background combination is found in an event,
all combinations are used.

The variables considered in the random forest algorithm are
the mass of the trijet system,
the mass of each dijet combination,
the angular separation and momenta of the jets in the trijet rest frame,
the  {\cPqb} tagging discriminator value of each jet,
and the quark-versus-gluon-jet
discriminator~\cite{CMS-PAS-JME-13-002} value of each jet.
To reduce correlations with the top quark \pt and thus
to prevent overtraining in this variable,
the \pt spectra of signal and background triplet combinations
are flattened through reweighting.
The random forest performance is improved by replacing
the kinematic variables in the laboratory frame with their
equivalents in the trijet rest frame,
and by sorting jets according to their momenta in the trijet rest frame so
that the highest (lowest) momentum jet is
most (least) likely to originate from a $\cPqb$ quark.

Trijet top quark candidates are selected by requiring
the random forest discriminator value to exceed 0.85.
This value is chosen based on optimization studies involving
the full limit-setting procedure described in Section~\ref{sec:results}.
If two or more selected trijets share one or more AK4 jets,
only the combination with the largest discriminator value is retained.

\begin{figure}[!th]
\centering
\includegraphics[width=\cmsFigWidth]{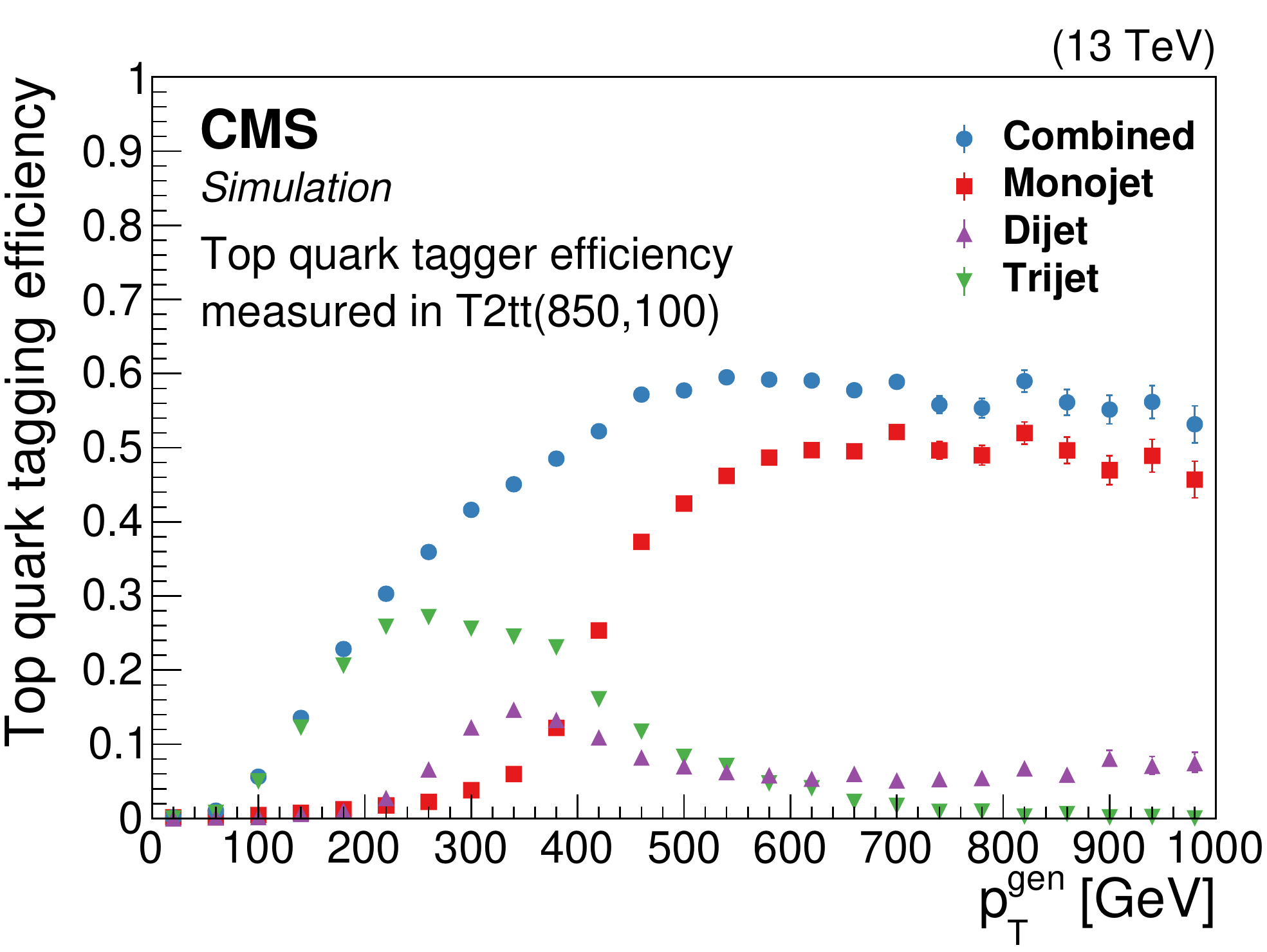}
\caption{
  Efficiency of the top quark tagger as a
  function of generator-level top quark \pt
  for the monojet (red boxes), dijet (magenta triangles),
  and trijet (green upside-down triangles) categories and for
  their combination (blue circles),
  as determined using T2tt signal events with a top squark mass
  of 850\GeV and an LSP mass of 100\GeV.
  The vertical bars indicate the statistical uncertainties.
  }
\label{fig:toptagger}
\end{figure}

All top quark candidates must have $\abs{\eta}<2.0$.
The final set consists of
the nonoverlapping candidates from the three reconstruction categories.
The total efficiency of the algorithm,
including a breakdown into the three categories,
is shown in Fig.~\ref{fig:toptagger}.
The efficiency is determined using T2tt signal events
with a top squark mass of 850\GeV and an LSP mass of 100\GeV,
based on the number of generator-level hadronically decaying
top quarks that are matched
to a reconstructed top quark candidate divided by the
total number of generator-level top quarks that decay hadronically.
Similar results are found using SM \ttbar events.
The matching between the generator-level and reconstructed
top quarks requires the overall reconstructed top quark to be
matched to the generator-level top quark within $\Delta R<0.4$.
The misidentification rate varies
between 15 and 22\% as a function of \ptmiss,
with an average of about 20\%,
as determined using simulated \znunujets events
after applying selection criteria similar to those used for the data
(Section~\ref{sec:selection}):
$\njets\geq4$,
$\nbjets\geq1$,
$\ptmiss>250\GeV$,
and no isolated electron or muon with $\pt > 10\GeV$.

Relative to Ref.~\cite{Khachatryan:2017rhw},
the top quark tagging algorithm has been improved by using
AK8 jets for the monojet and dijet categories,
rather than strictly AK4 jets,
and through implementation of the random forest tree
for the trijet category.
These improvements provide a factor of two reduction in
the top quark misidentification rate
while maintaining a similar efficiency.

\section{Event selection and search regions}
\label{sec:selection}

Our study is an inclusive search for events containing
\ptmiss and reconstructed top quarks.
The selection criteria are intended, in general, to be nonrestrictive,
while still providing high trigger efficiency and
sensitivity to a wide variety of new-physics scenarios.
All events must satisfy filters designed to remove
detector- and beam-related noise.
The events are subjected to the lepton, isolated-track,
and charged-hadron vetoes of Section~\ref{sec:vetoes}.
To improve the rejection of background,
the two tight AK4 jets with highest \pt must have $\pt>50\GeV$.
Events are required to have $\njets\geq4$, $\nbjets\ge 1$,
$\ntops\ge 1$, $\ptmiss>250\GeV$, and $\HT>300\GeV$.

The QCD multijet background mostly arises when the \pt of one
of the highest \pt jets is undermeasured,
causing \ptvecmiss to be aligned with that jet.
This undermeasurement can occur because of jet misreconstruction or,
in the case of semileptonic $\cPqb$ or $\cPqc$ quark decays,
an undetected neutrino.
To reduce this background,
requirements are placed on the azimuthal angle between \ptvecmiss
and the three loose AK4 jets with highest~\pt,
denoted $j_1$, $j_2$, and $j_3$ in order of decreasing~\pt.
Specifically,
we require
$\Delta\phi(\ptvecmiss, j_{1})>0.5$,
$\Delta\phi(\ptvecmiss, j_{2})>0.5$,
and
$\Delta\phi(\ptvecmiss, j_{3})>0.3$.

The \MTTwo variable~\cite{Lester:1999tx,Barr:2003rg,Khachatryan:2017rhw}
is used to reduce background from \ttbar events.
This variable is designed to provide an estimate of the transverse mass
of pair-produced heavy objects
that decay to both visible and undetected particles.
It has a kinematic upper limit at the mass of
the heavy object undergoing decay.
Thus the upper limit for SM \ttbar events is $m_\cPqt$,
while the upper limit for TeV-scale squarks and gluinos is much larger.
If there are two tagged top quarks in an event,
\MTTwo is calculated using the pair of tagged top quarks and \ptvecmiss.
If there are more than two tagged top quarks,
we compute \MTTwo for all combinations and choose the
combination with the smallest \MTTwo.
If there is only one tagged top quark,
we construct a proxy for the other top quark
using the highest \pt $\cPqb$ tagged jet as a seed.
If a $\cPqb$ tagged jet is not available,
because there is only one $\cPqb$ tagged jet in the event and
it is part of the reconstructed top quark,
the highest \pt jet is used as the seed.
The seed jet is combined with a loose AK4 jet to define the top quark proxy
if the resulting pair of jets has a mass between 50 and 220\GeV
and if the two jets appear within $\Delta R=1.5$ of each other;
otherwise the seed jet by itself is used as the top quark proxy.
The proxy is combined with the tagged top quark and \ptvecmiss
to determine~\MTTwo.
Irrespective of the number of
tagged top quarks,
we require $\MTTwo>200\GeV$.

\begin{figure*}[htb]
  \centering
    \includegraphics[width=0.32\linewidth]{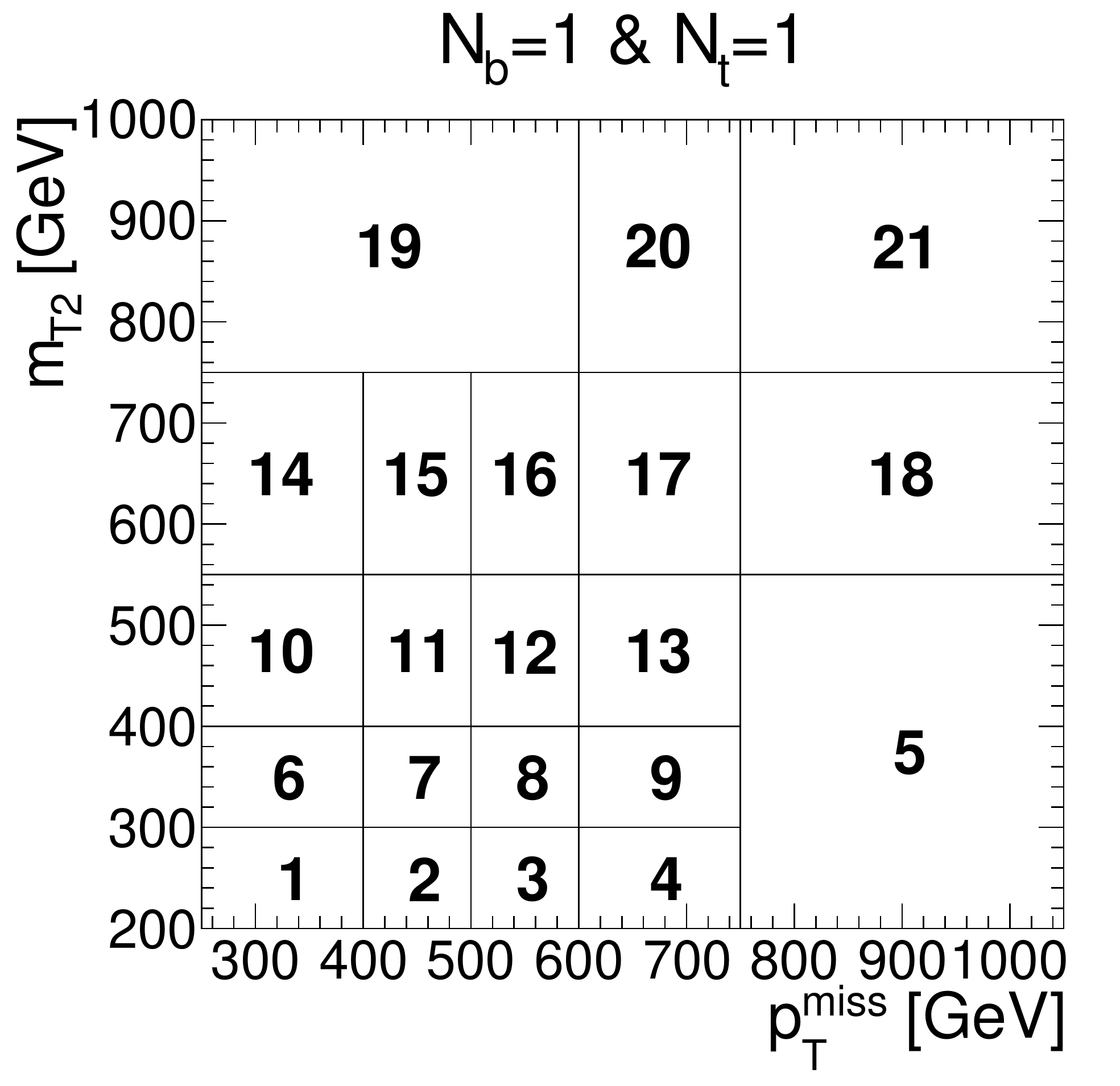}
    \includegraphics[width=0.32\linewidth]{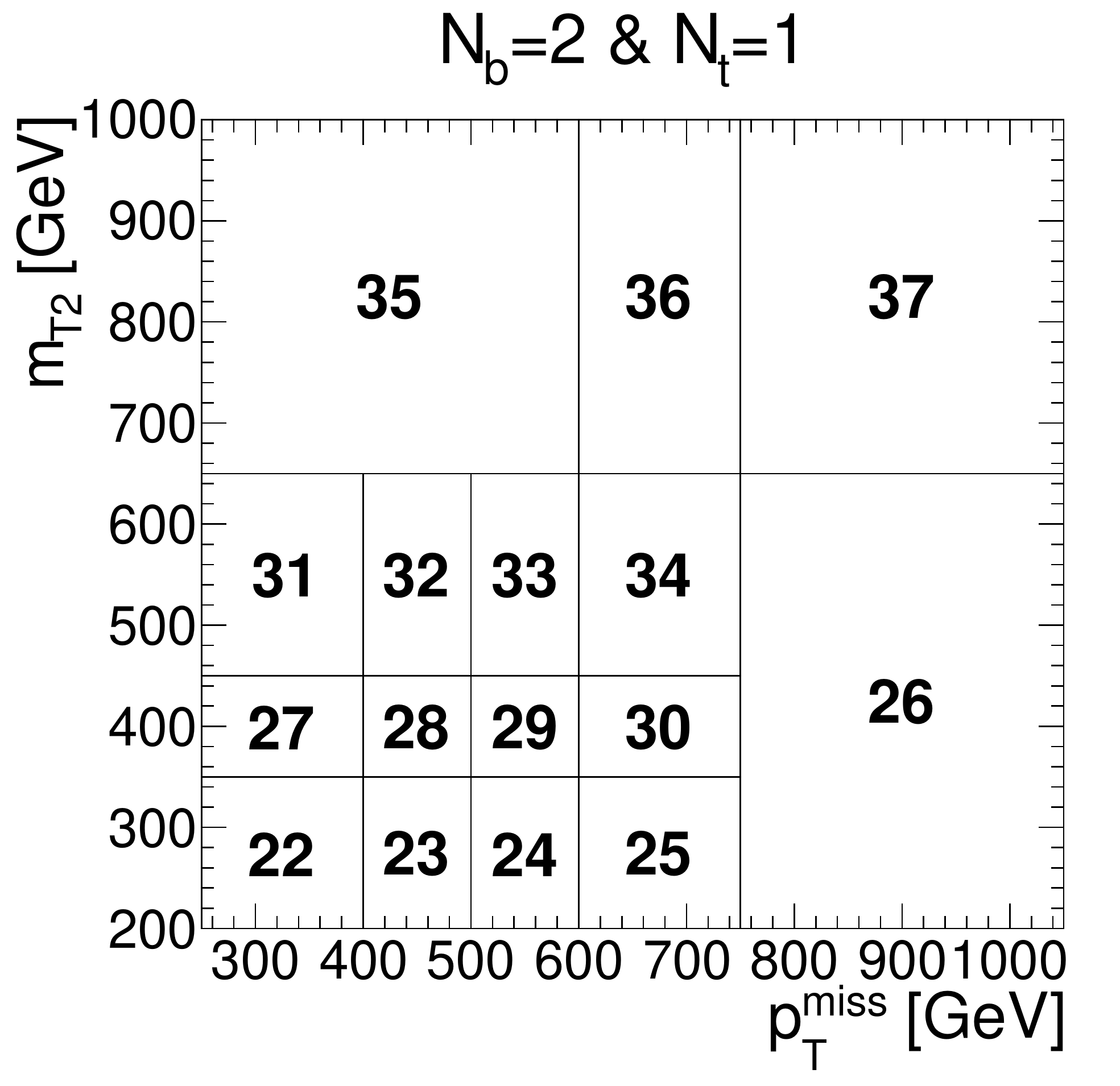}
    \includegraphics[width=0.32\linewidth]{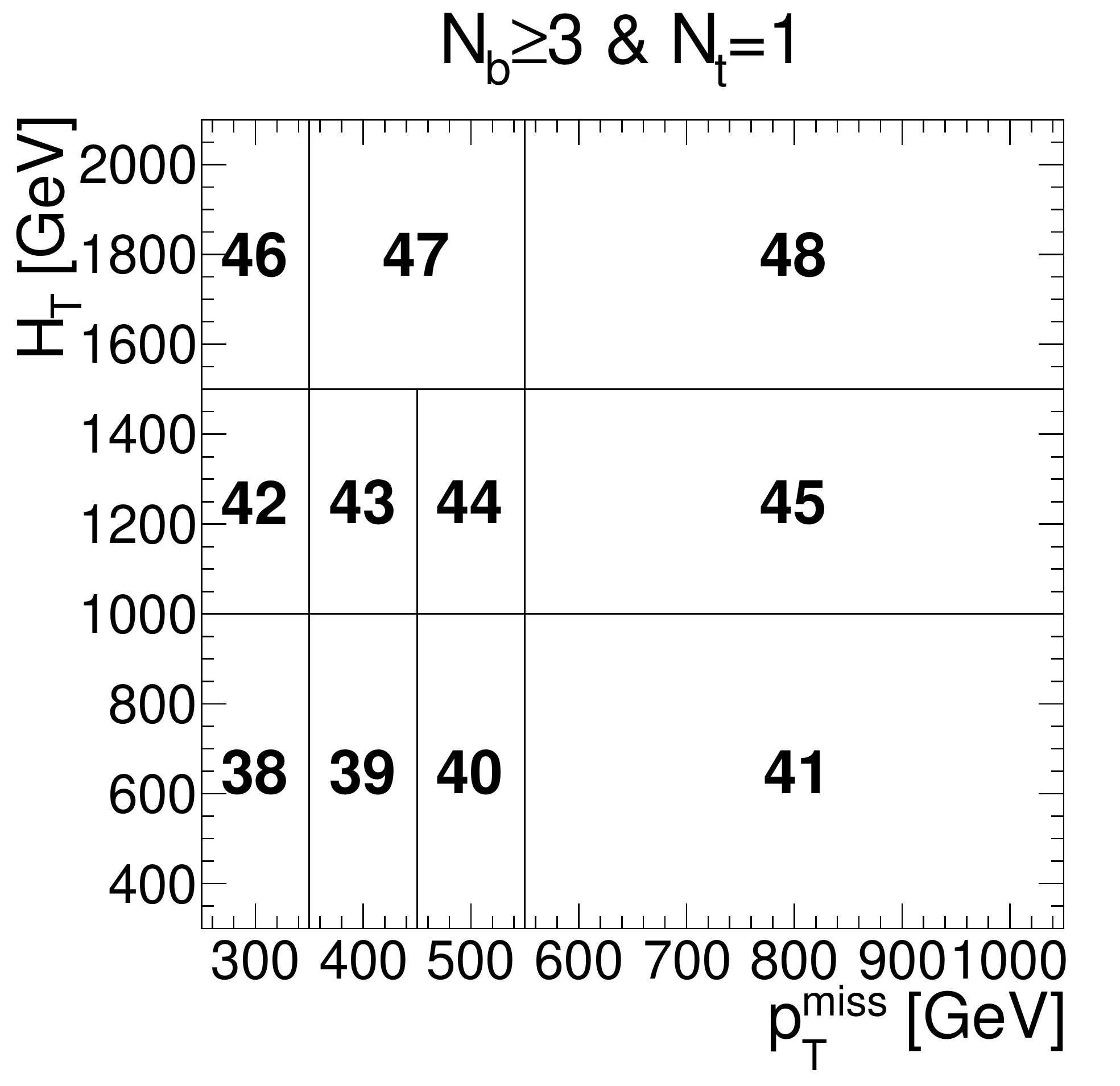} \\
    \includegraphics[width=0.32\linewidth]{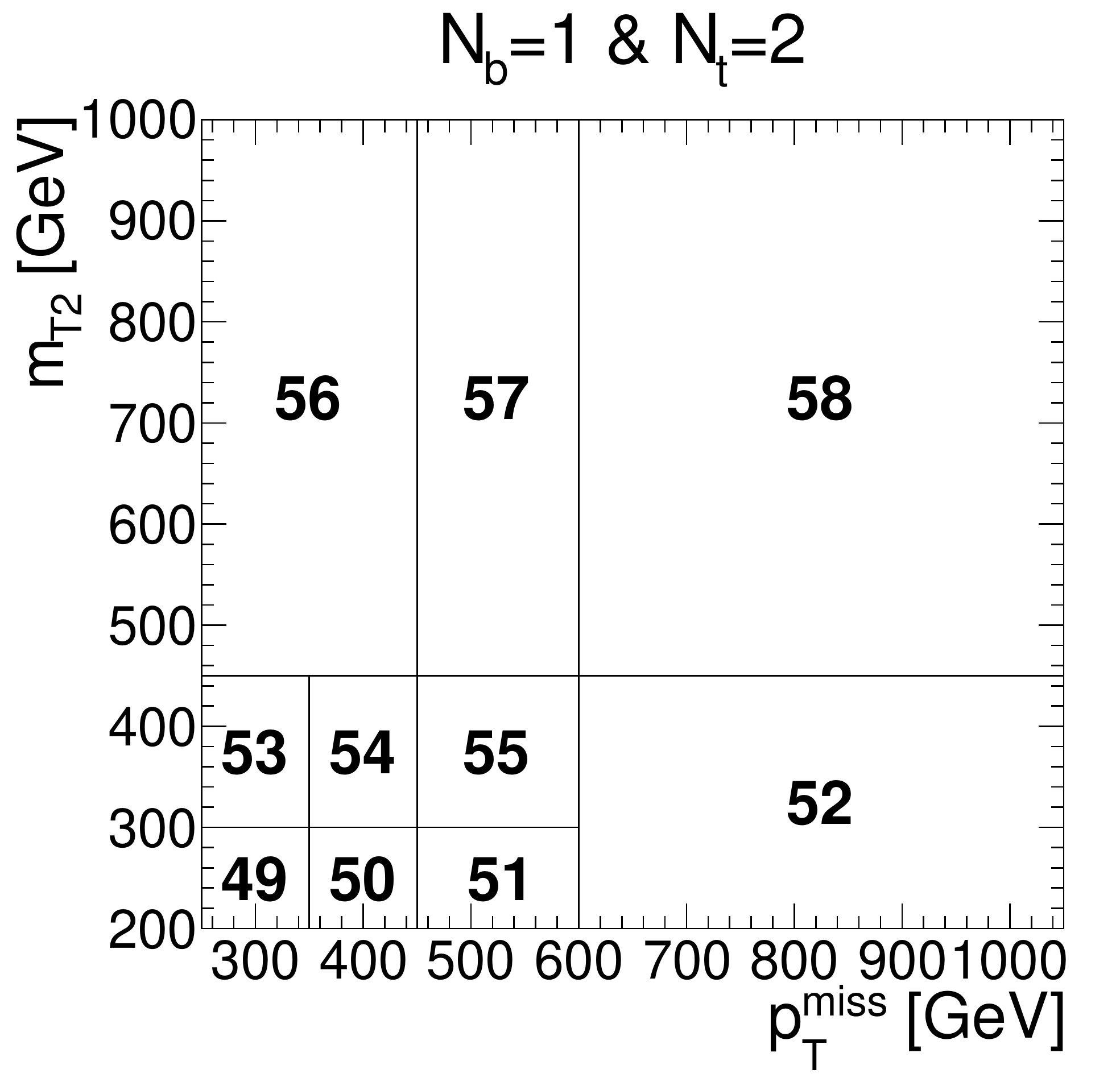}
    \includegraphics[width=0.32\linewidth]{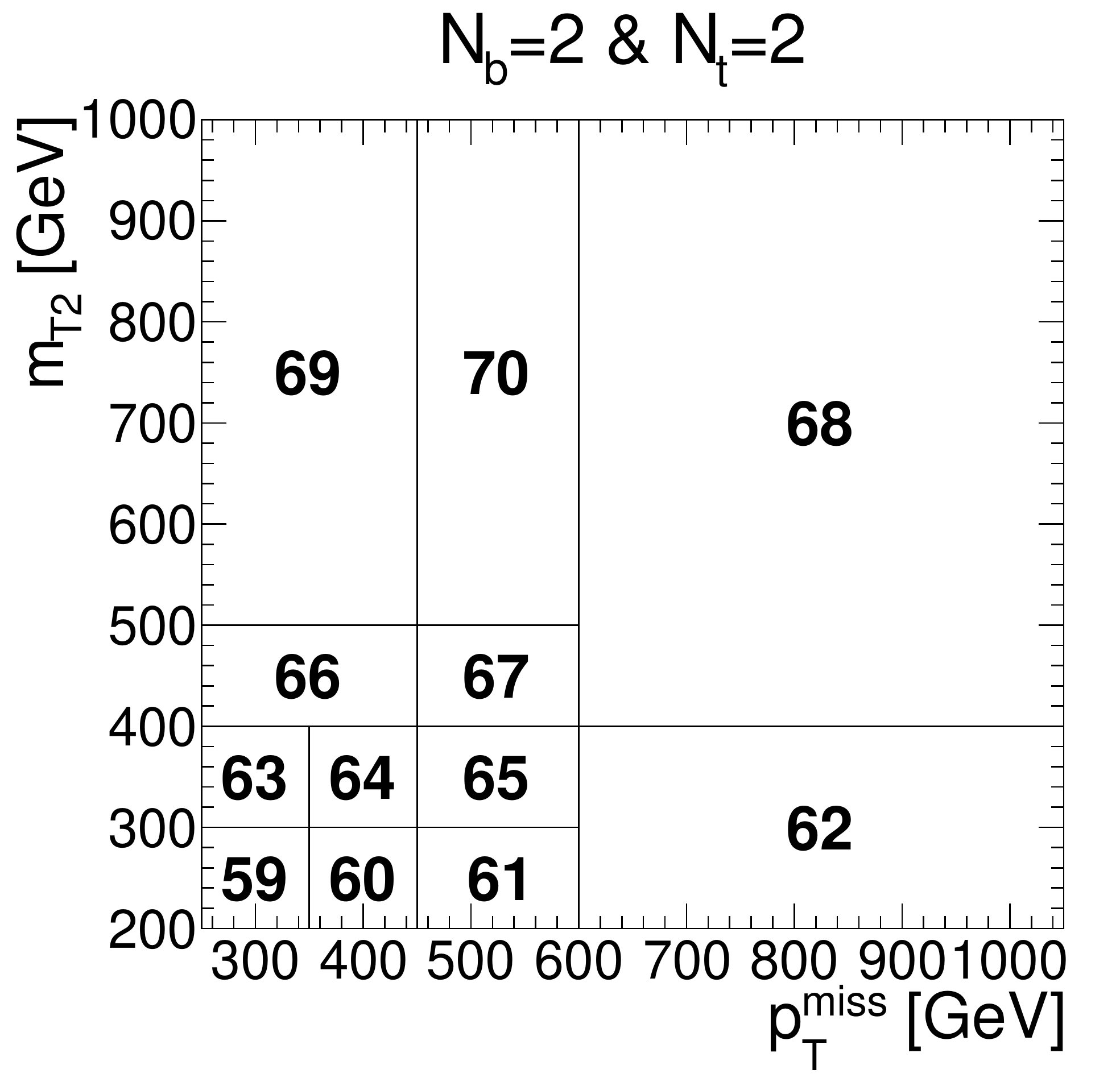}
    \includegraphics[width=0.32\linewidth]{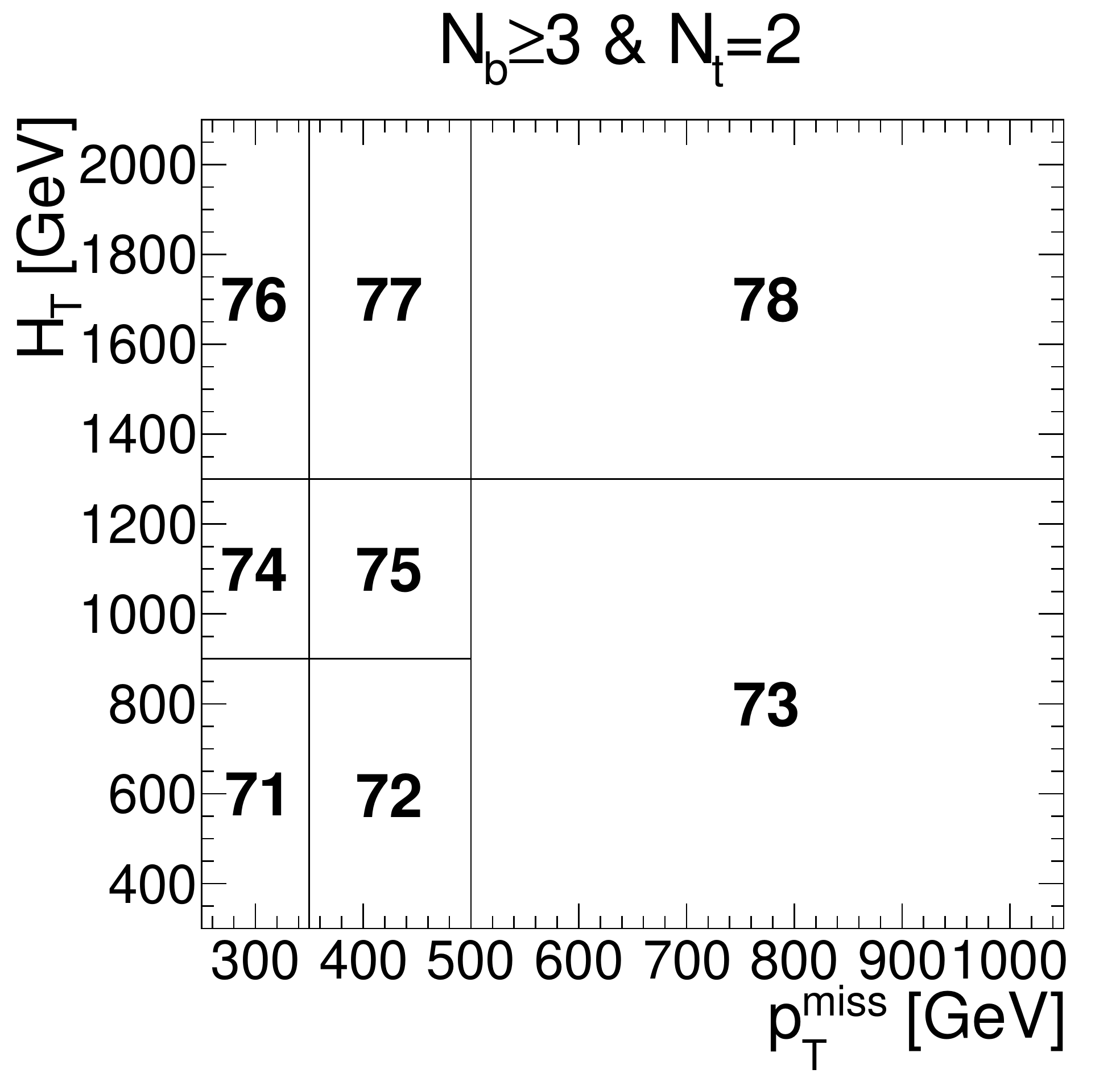} \\
    \includegraphics[width=0.32\linewidth]{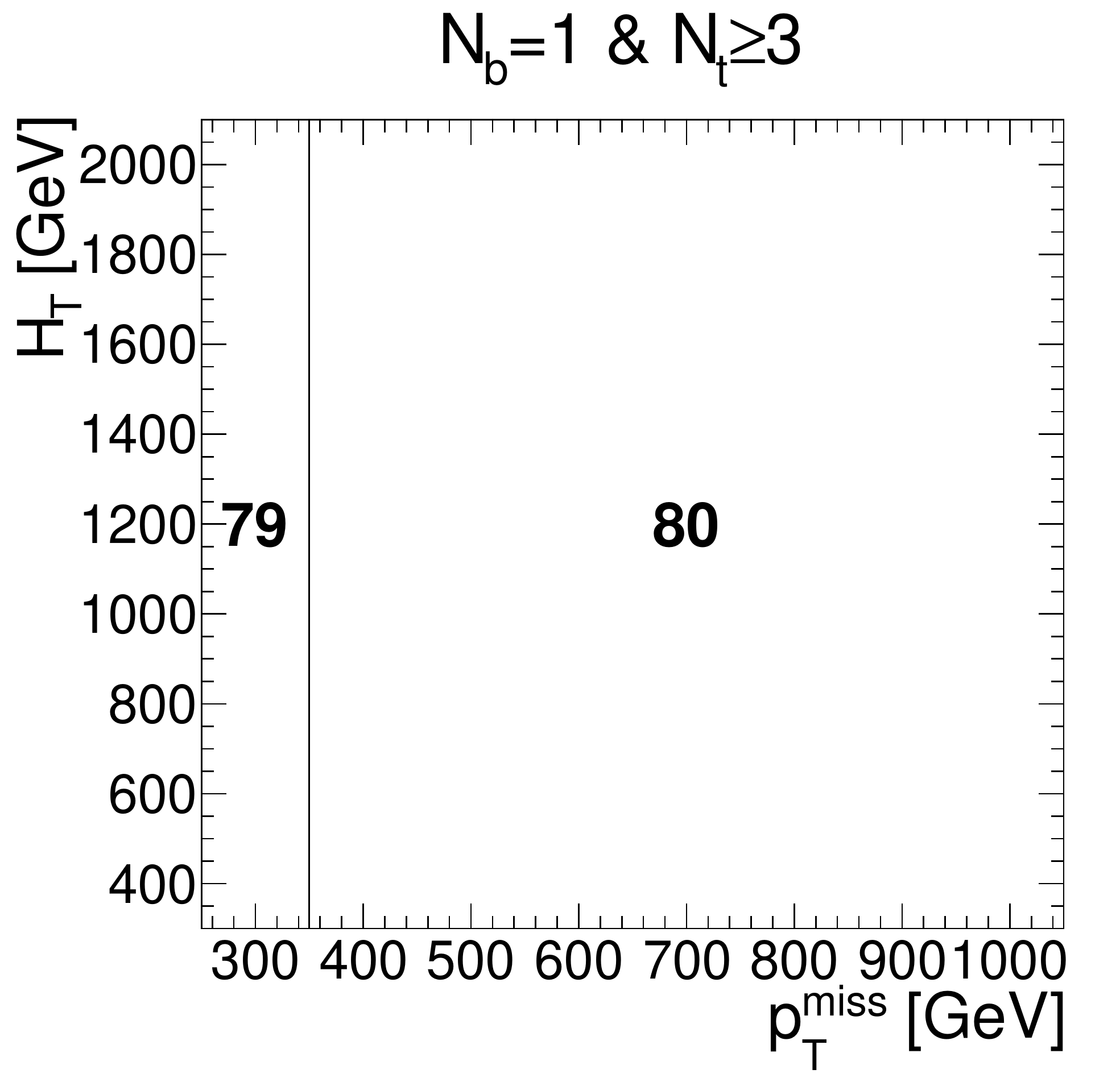}
    \includegraphics[width=0.32\linewidth]{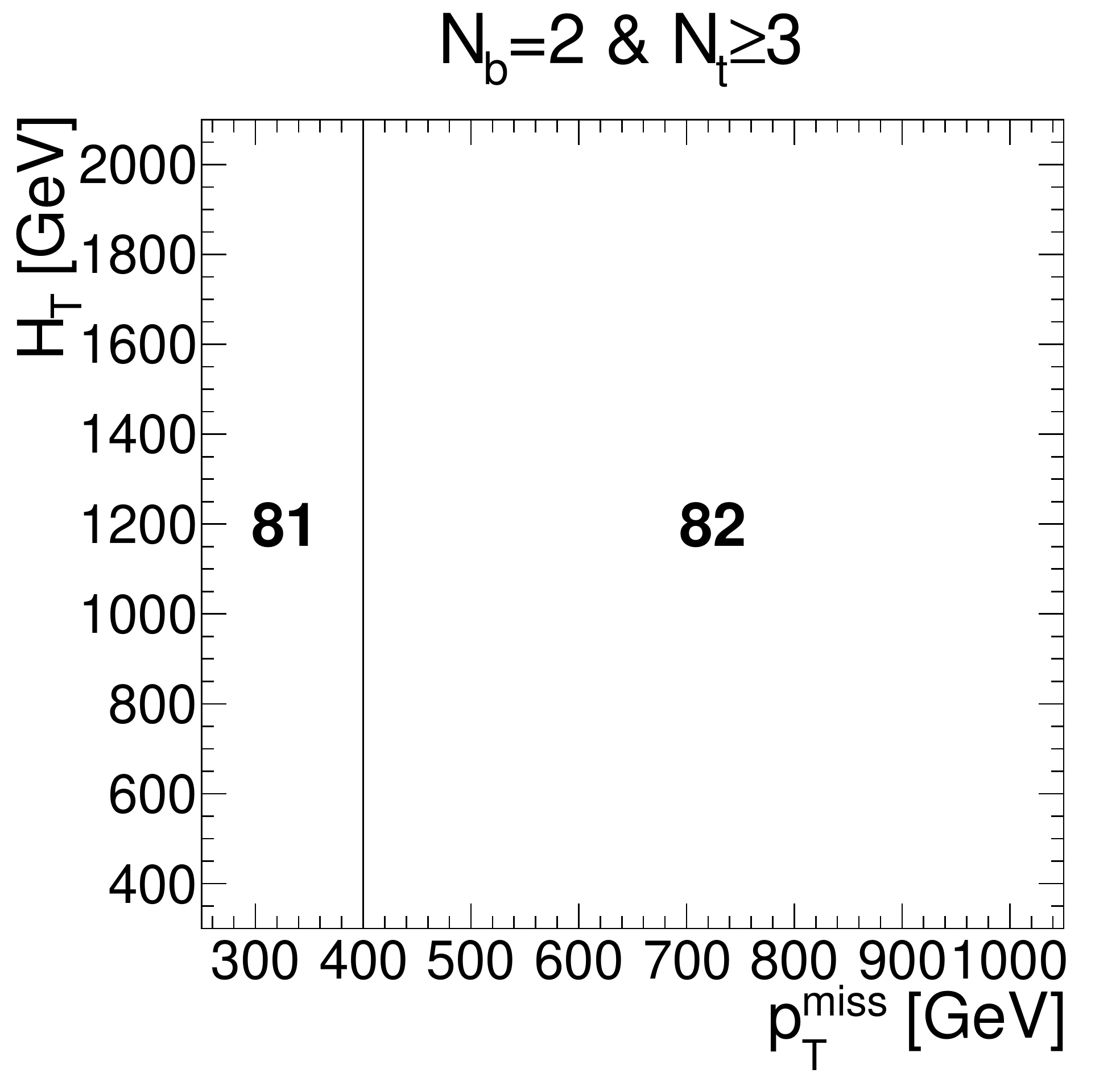}
    \includegraphics[width=0.32\linewidth]{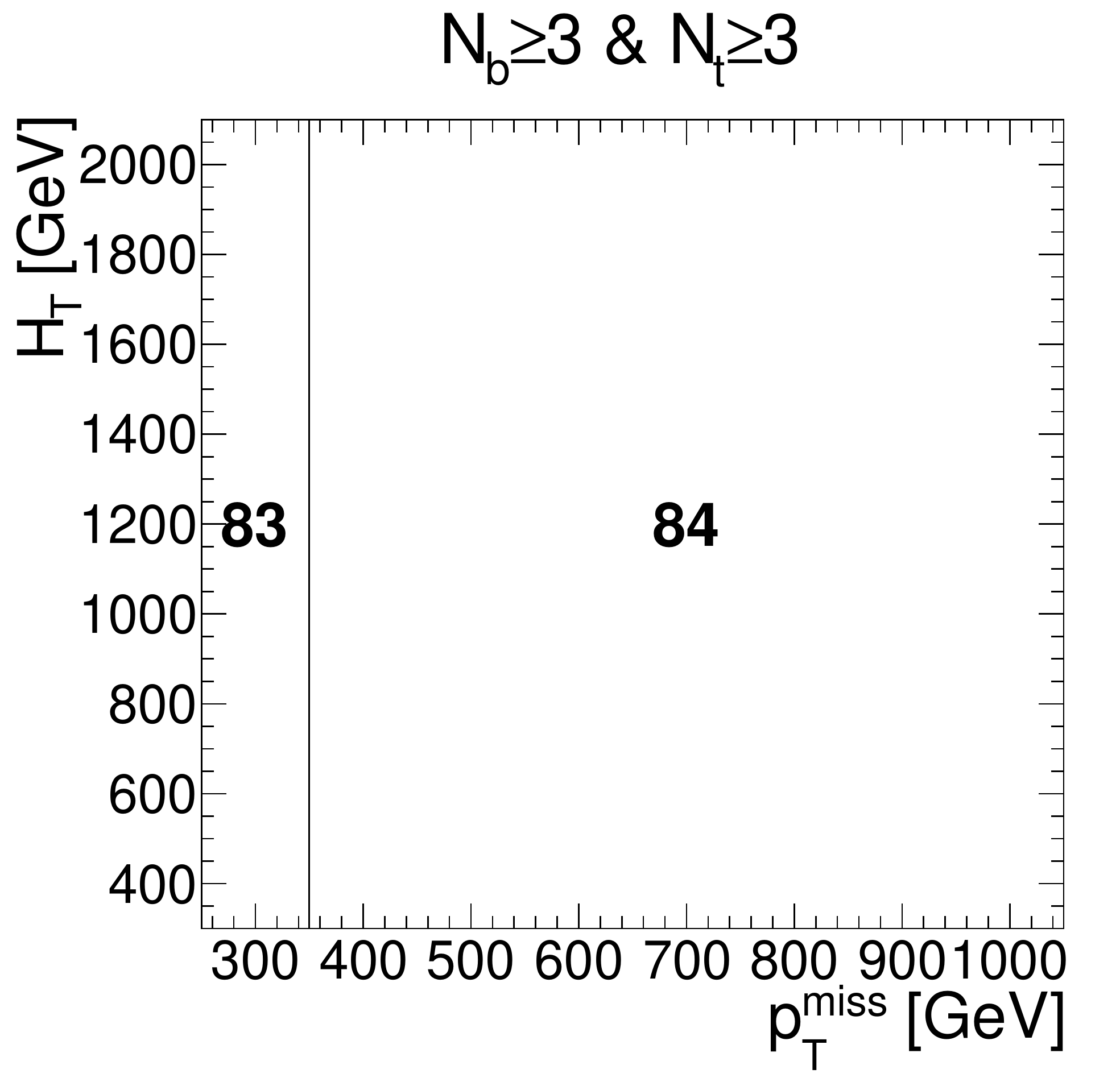} \\
    \caption{Search region definitions in the kinematic variables.
The highest \ptmiss, \MTTwo, and \HT regions are open-ended, \eg,
$\ptmiss>750\GeV$ and $\MTTwo > 750\GeV$ for search region~21.
}
\label{fig:binning}
\end{figure*}

The search is performed in 84 nonoverlapping search regions.
Regions with $\nbjets\leq2$ and $\ntops\leq2$ use
\nbjets, \ntops, \ptmiss, and \MTTwo
as the binned search variables.
Regions with $\nbjets\geq3$ or $\ntops\geq3$
use \nbjets, \ntops, \ptmiss, and \HT.
The reason \HT is used for these latter regions,
and not \MTTwo,
is that in events with many jets,
the jets from the decay of a particular heavy object
may not always be correctly associated with that object,
causing the distribution of \MTTwo
to be broad and relatively flat.
We find that \HT provides better discrimination between signal
and background for $\nbjets\geq3$ or $\ntops\geq3$.
The 84 regions in \MTTwo versus \ptmiss or in \HT versus \ptmiss
are illustrated in Fig.~\ref{fig:binning}.
The boundaries between the regions were determined through sensitivity studies.

To simplify use of our data by others,
we also define 10 aggregate search regions,
specified in Table~\ref{tab:aggBinDescrption}.
The aggregate regions are nonexclusive
and are intended to be considered independently.
The first four aggregate regions represent topologies of general interest.
The fifth and sixth are sensitive to
direct top squark pair production.
The seventh region targets
the large $\Delta m(\gluino,\lsp)$ region of T5ttcc-like models,
while the final three target events with a large number of top quarks
such as are produced in the T1tttt and T5tttt models.

\begin{table*}[htb]
\centering
  \topcaption{Definition of the aggregate search regions.}
\label{tab:aggBinDescrption}
\renewcommand{\arraystretch}{1.15}
\begin{scotch}{cccccl}
Region & \ntops & \nbjets & \MTTwo [\GeVns{}] & \ptmiss [\GeVns{}] & \multicolumn{1}{c}{Motivation} \\
\hline
1  & $\geq$1     & $\geq$1      & $\geq$200       & $\geq$250        & Events satisfying selection criteria      \\
2  & $\geq$2     & $\geq$2      & $\geq$200       & $\geq$250        & Events with $\ntops \ge 2$ and $\nbjets \ge 2$  \\
3  & $\geq$3     & $\geq$1      & $\geq$200       & $\geq$250        & Events with $\ntops \ge 3$ and $\nbjets \ge 1$  \\
4  & $\geq$3     & $\geq$3      & $\geq$200       & $\geq$250        & T5tttt; small $\Delta m(\gluino,\lsp)$ and $\mlsp<m_{\cPqt}$     \\
5  & $\geq$2     & $\geq$1      & $\geq$200       & $\geq$400        & T2tt; small $\Delta m(\sTop,\lsp)$      \\
6  & $\geq$1     & $\geq$2      & $\geq$600       & $\geq$400        & T2tt; large $\Delta m(\sTop,\lsp)$      \\
\hline
Region & \ntops & \nbjets & \HT [\GeVns{}] & \met [\GeVns{}] & Motivation \\
\hline
7  & $\geq$1     & $\geq$2      & $\geq$1400      & $\geq$500        & T1ttbb \& T5ttcc; large $\Delta m(\gluino,\lsp)$    \\
8  & $\geq$2     & $\geq$3      & $\geq$600       & $\geq$350        & T1tttt; small $\Delta m(\gluino,\lsp)$    \\
9  & $\geq$2     & $\geq$3      & $\geq$300       & $\geq$500        & T1/T5tttt \& T1ttbb; intermediate $\Delta m(\gluino,\lsp)$    \\
10 & $\geq$2     & $\geq$3      & $\geq$1300      & $\geq$500        & T1/T5tttt; large $\Delta m(\gluino,\lsp)$    \\
\end{scotch}
\end{table*}

\section{Background estimation}
\label{sec:backgroundestimation}

We next discuss the evaluation of the SM background.
A change relative to Ref.~\cite{Khachatryan:2017rhw}
is that we now use a translation factor method,
as described in Section~\ref{sec:TopW},
to evaluate the background from \ttbar, single top quark, and \wjets events.
In Ref.~\cite{Khachatryan:2017rhw} we
rather used \tauh response templates and
separately evaluated terms constructed from the electron and muon acceptance,
isolation efficiency,
and reconstruction-and-identification efficiency
to evaluate this background.
The reason for the change is to simplify
the modeling of variables for the
AK8 jets and for the random forest decision tree
now used in the top quark tagging algorithm.
Another change is that the ``loose'' dimuon control sample
described in Section~\ref{sec:Zinv}
is selected using more restrictive requirements,
as is allowed by the larger data sample now available,
leading to reduced systematic uncertainties.

\subsection{Background from \texorpdfstring{\ttbar}{ttbar}, single top quark, and \texorpdfstring{\wjets}{W+jets} events}
\label{sec:TopW}

The largest background,
accounting for about 70\% of the total background
integrated over the 84 search regions,
is due to \ttbar, single top quark, and {\wjets} events
with a leptonically decaying {\PW} boson.
This background arises in one of two distinct ways.
First, if the $\PW$ boson decays to a $\tau$ lepton that decays hadronically,
the $\tau$ lepton can be reconstructed as a jet and the event can escape
the vetoes of Section~\ref{sec:vetoes}.
Second, if the $\PW$ boson decays to an electron or muon
(including from the decay of a $\tau$ lepton)
that is not reconstructed or identified,
is not isolated, or lies outside the acceptance of the analysis,
the event can escape the vetoes.
These two possibilities are referred to as
the \tauh and lost-lepton backgrounds, respectively.
They are evaluated, together,
using a single-lepton data control sample (CS)
collected using the same trigger that is used to collect signal events.
The CS events must satisfy the same criteria as the data
except for the vetoes of Section~\ref{sec:vetoes},
which are replaced by a requirement that there be exactly
one isolated electron or muon candidate based on
the isolation criteria of Section~\ref{sec:vetoes}.
To reduce potential contributions from signal processes,
CS events must have $\mt<100\GeV$.

The predicted summed number of \tauh and lost-lepton events
in a search region is given by
the number of single-electron or single-muon
events in the corresponding region of the CS,
multiplied by a translation factor from simulation.
Predictions from the single-electron and single-muon samples
are determined separately and used as independent constraints
in the likelihood fit described in Section~\ref{sec:results}.
The translation factor is given by
the ratio of the summed number of simulated \tauh and lost-lepton events
in the search region to the number of simulated single-electron or
single-muon events in the corresponding CS region.

\begin{figure}[tbp]
 \centering
 \includegraphics[width=0.48\textwidth]{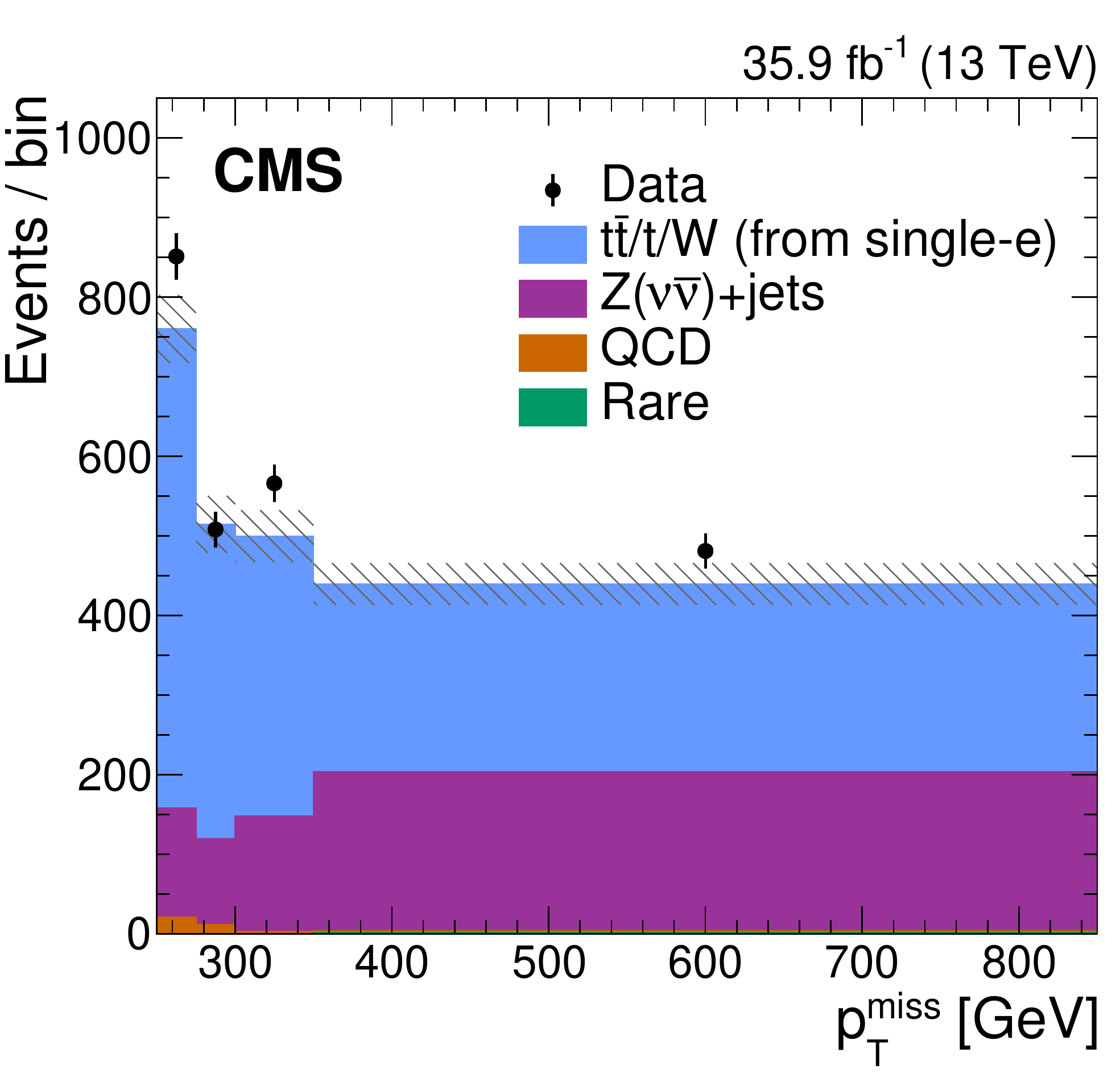}
 \includegraphics[width=0.48\textwidth]{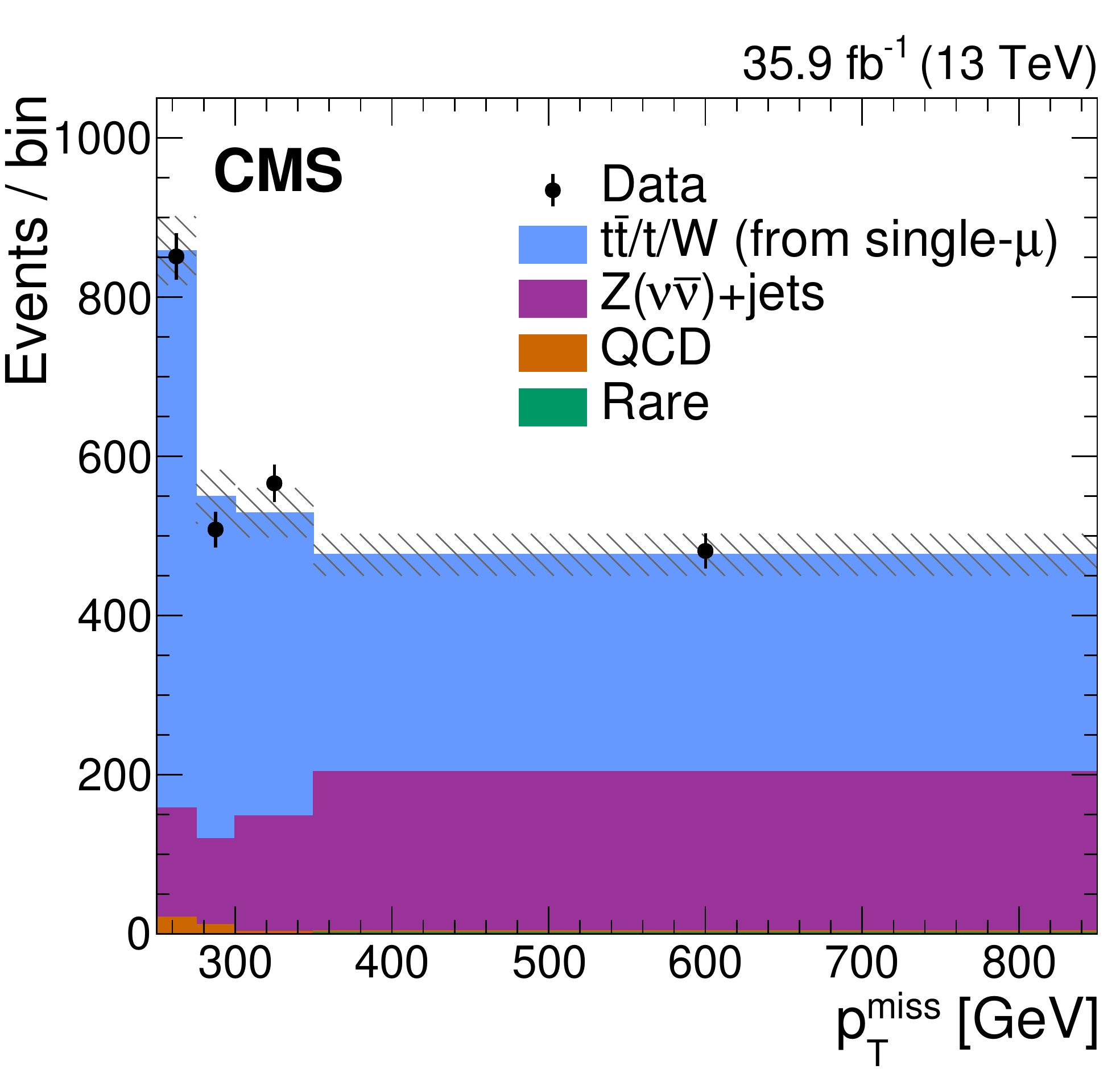}
\caption{
Distribution of \ptmiss in the sideband data sample
in comparison to predictions for SM processes.
The prediction for \ttbar, single top quark, and {\wjets} events
is obtained using translation factors applied to
a single-electron control sample (\cmsLeft)
or to a single-muon control sample (\cmsRight).
The hatched bands indicate the statistical uncertainties
in the total SM prediction.
Note that the data and the predictions for all backgrounds
except that for \ttbar, single top quark, and {\wjets} events
are identical between the left and right plots.
}
\label{fig:val_0t2b}
\end{figure}

The method is tested using an orthogonal data sample,
referred to as the ``sideband'' (SB),
selected using the same criteria as are applied to the data except
with $\ntops=0$, $\nbjets\geq2$,
and $\Delta\phi(\metv, j_{1,2,3,4})>0.5$,
where the last two requirements reduce contributions from \znunujets
and QCD multijet events.
The SB,
which is enhanced in events with semileptonic top quark decays,
is divided into four intervals of \ptmiss.
The contribution of \tauh and lost-lepton events to
the intervals is determined
in an analogous manner to that described above for the search regions,
namely by multiplying the number of events in
the corresponding interval of the single-electron
or single-muon CS by a translation factor from simulation,
defined analogously to the translation factors of the standard analysis.
The contributions of \zjets, QCD multijet, and rare events to the SB
are taken directly from simulation.
Figure~\ref{fig:val_0t2b} shows the \ptmiss distribution in the SB
in comparison to the SM prediction.
The histogram labeled ``$\ttbar/\cPqt/{\PW}$''
shows the predicted contribution from \tauh and lost-lepton events.
The total SM prediction is seen to agree with the data
within the uncertainties,
providing a validation for the translation factor procedure.

Systematic uncertainties in the
prediction for the \ttbar, single top quark, and {\wjets} background
are evaluated from the following sources,
based on the uncertainties in the respective quantities:
the statistical uncertainty in the translation factors
(1--40\% depending on the search region),
the lepton reconstruction and isolation efficiency (7--43\%),
the jet and \ptmiss energy scale and resolution (up to 64\%),
the ISR modeling (up to 13\%),
the PDFs (up to 32\%),
and the $\cPqb$ jet tagging efficiency (1\%).

As a cross-check,
the lost-lepton background is evaluated
using a complementary procedure,
described in Ref.~\cite{Khachatryan:2017rhw},
based on the single-lepton CS described above
and on factors obtained for each search region
from \ttbar, single top quark, and {\wjets} simulation
that account for the acceptance,
the isolation efficiency,
and the reconstruction-and-identification efficiency.
The lost-lepton background evaluated with this approach
is consistent with that obtained from the translation factor method.

\subsection{\texorpdfstring{Background from \znunujets events}{Background from Znunu events}}
\label{sec:Zinv}

The background from \znunujets events is evaluated
using simulated \znunujets events that
satisfy the search region selection criteria.
Two corrections,
derived from an event sample enhanced in
DY($\cPZ\to\mu\mu$)+jets production,
are applied to account for differences between data and simulation.
The trigger to select the DY+jets events requires that
there be at least one muon with $\pt>50\GeV$,
while the offline selection requires
two oppositely charged muons with
a dimuon invariant mass between 81 and 101\GeV,
and the highest (second-highest) \pt muon in the event
to have $\pt>50\,(20)\GeV$.
The dimuon system is removed from the events to emulate \ptvecmiss
in {\znunujets} events.

The first correction,
which accounts for the \njets distribution,
is based on a ``loose'' dimuon control sample
selected by imposing,
on the DY-enhanced event sample described in the previous paragraph,
the same requirements on
$\Delta\phi(\metv, j_{1,2,3})$, \HT, and \ntops as are applied
to signal candidate events,
but with the less stringent requirement $\ptmiss>100\gev$
and with no requirement on \nbjets.
The correction is determined as a function of \njets
as the ratio of the number of events in the loose control sample,
with non-DY events subtracted using simulation,
to the number of events in a similarly selected sample of simulated DY events.
The corrections are applied to the \znunujets simulation
as weights based on the value of \njets.

The second correction adjusts the overall normalization
of the simulated \znunujets sample.
It is derived from a ``tight'' dimuon data control sample
selected by applying,
to the DY-enhanced event sample described
in the first paragraph of this section,
the same requirements as are applied to signal events except,
of the vetoes described in Section~\ref{sec:vetoes},
only the veto on isolated electrons is applied,
and there is no requirement on \nbjets.
The correction is given by the ratio of the number of events
in the tight control sample,
with non-DY backgrounds subtracted using simulation,
to the number of events in a sample of simulated DY events
selected with the same criteria.

\begin{figure}[tbp]
 \centering
 \includegraphics[width=0.48\textwidth]{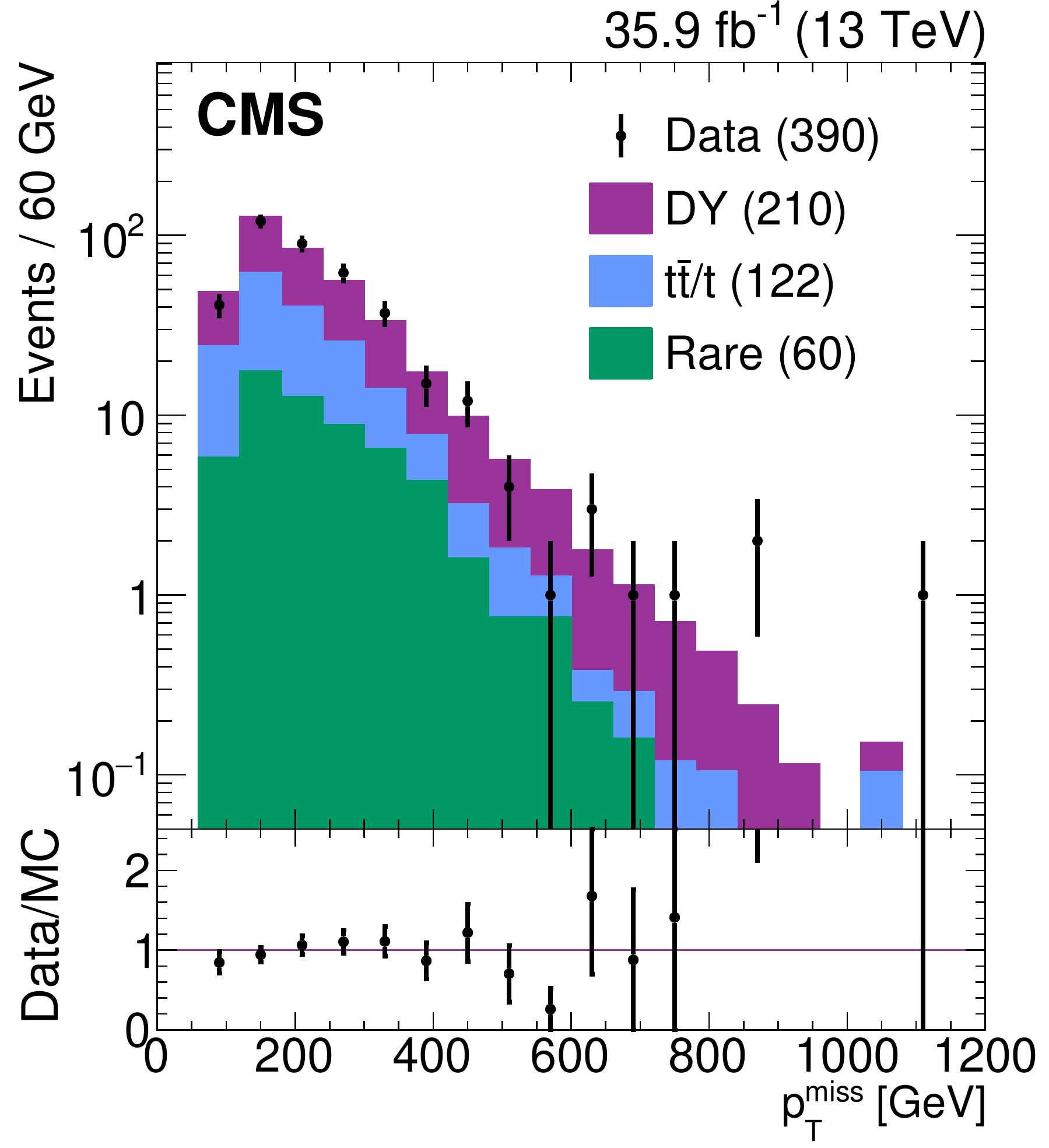}
 \includegraphics[width=0.48\textwidth]{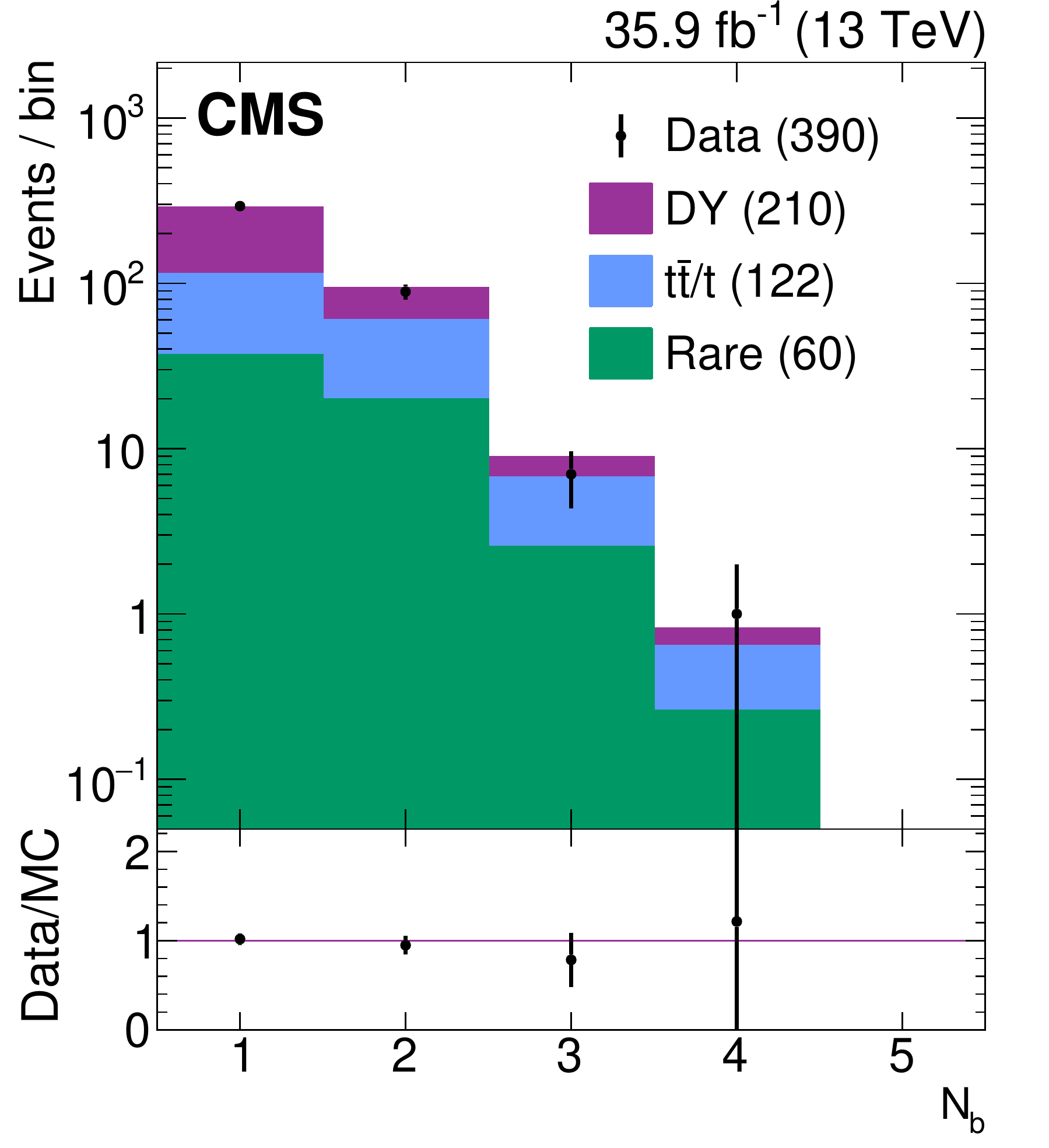}
\caption{
The \ptmiss (\cmsLeft) and \nbjets (\cmsRight) distributions
of data and simulation in the loose
dimuon control sample
after applying a correction,
as described in the text,
to account for differences between the data and
simulation for the \njets distribution.
The lower panels show the ratio between data and simulation.
Only statistical uncertainties are shown.
The values in parentheses indicate the integrated yields
for each component.
}
\label{fig:MethodAlpha_zinv_loose}
\end{figure}

Systematic uncertainties in the prediction
for the {\znunujets} background
are derived from the shape differences between data and simulation
in the loose dimuon control sample as a function of
$\nbjets$, $\ntops$, \ptmiss, \MTTwo, and \HT after the
first correction described above has been applied.
As examples,
the post-correction comparisons between data and simulation
for the \ptmiss and \nbjets distributions
are shown in Fig.~\ref{fig:MethodAlpha_zinv_loose}.
The shift in the central value between the data and simulation in the
distributions is used to define an additional uncertainty,
which varies between 14 and 44\% depending on the search region.
The statistical uncertainty in the \njets shape correction (1--46\%)
and in the overall normalization correction (7.6\%)
are also taken as systematic uncertainties.
Additional systematic uncertainties account for the
jet and \ptmiss energy scales (1--71\%),
the $\cPqb$ tagging efficiency (1--23\%),
the PDFs and the renormalization and factorization scales (1--48\%),
the statistical uncertainty in the simulation
(1--81\%, with the results for a few search regions as high as 100\%),
and the trigger (up to 14\%).

\subsection{Background from multijet events}
\label{sec:QCD}

The background from QCD multijet events is evaluated similarly
to the background from \ttbar, single top quark, and {\wjets} events.
A QCD data control sample is defined using the same trigger
and selection criteria as are used to select signal events
but with the less restrictive condition $\ptmiss>200\gev$ and with
the selection criteria on $\Delta\phi(\metv, j_{1,2,3})$ inverted.
This yields a signal-depleted control sample dominated by QCD multijet events.
The predicted number of QCD multijet events in each of the 84
search regions is given by the yield in the corresponding
region of the QCD control sample,
after contributions from non-QCD SM processes
have been subtracted using simulation,
multiplied by a translation factor derived from simulated QCD multijet events.
The translation factors are applied
as a function of \ptmiss and \MTTwo for \nbjets and $\ntops\le2$,
and as a function of \ptmiss for \nbjets or $\ntops\geq3$,
and are normalized to data
in the $200 < \ptmiss < 250 \gev$ region of the QCD control sample.

A systematic uncertainty in the QCD multijet prediction
for each search region
is evaluated as the difference between the event yield
obtained directly from the QCD multijet simulation for that region
and the prediction
obtained by applying the background prediction procedure
to simulated QCD multijet samples (30--500\%).
Additional sources of uncertainty are from the statistical
uncertainty in the translation factors (30--300\%) and
the subtraction of the non-QCD-multijet SM contributions
to the QCD control sample (2--50\%).

\subsection{Background from rare processes}
\label{sec:rare}

Background from rare events
forms only a small fraction of the total background
and has only a small effect on the final result.
Estimates of the rates of rare background processes
are taken directly from simulation.
The largest component of this background
is from $\ttbar\cPZ$ production.
To validate the $\ttbar\cPZ$ cross section in the simulation,
a three-lepton control sample is selected.
The yields of events in this sample between simulation
and data are found to agree within the statistical uncertainty of 30\%,
which is taken as the systematic uncertainty in the
$\ttbar\cPZ$ background estimate.

\section{Results and interpretation}
\label{sec:results}

\begin{figure*}[htb]
  \centering
  \includegraphics[width=\textwidth]{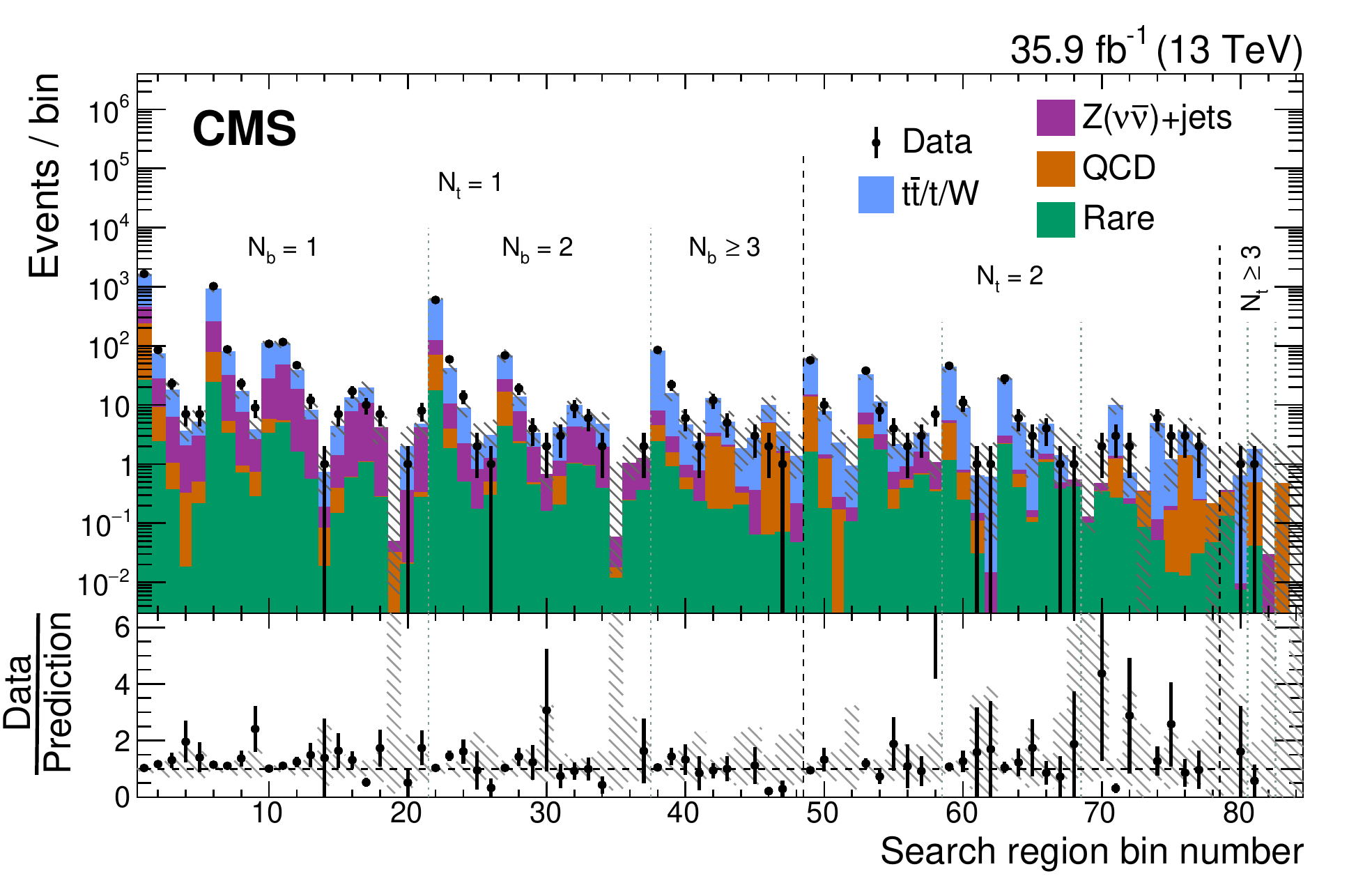}
  \caption{
  Observed event yields (black points)
  and prefit SM background predictions (filled solid areas) for the 84 search regions,
  where ``prefit'' means there is no constraint from the likelihood fit.
  The lower panel shows the ratio of the data to the total
  background prediction.
  The hatched bands correspond to the total uncertainty in the
  background prediction.
  }
  \label{fig:baseline_SR}
\end{figure*}

\begin{figure}[htb]
  \centering
  \includegraphics[width=\cmsFigWidth]{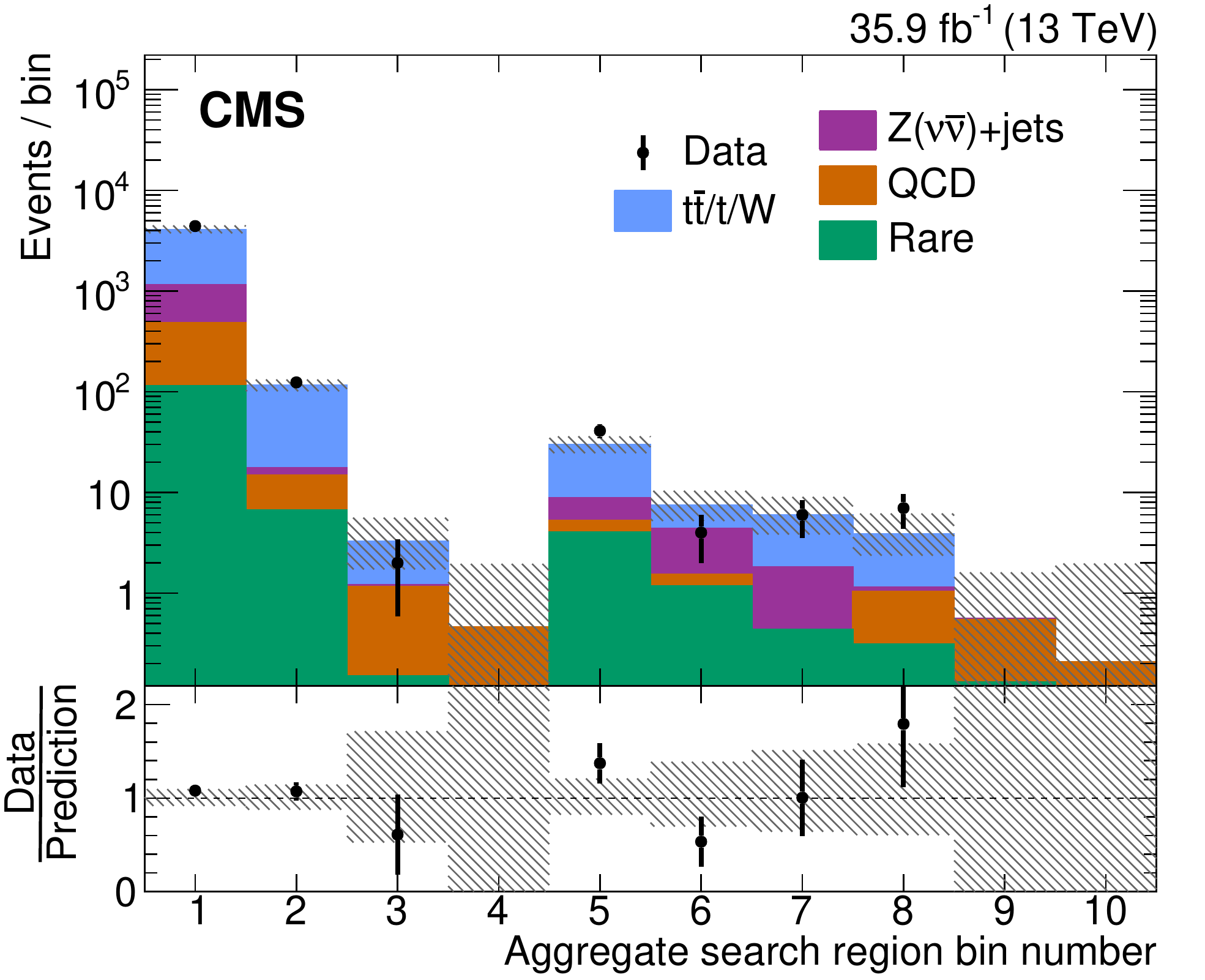}
  \caption{
  Observed event yields (black points)
  and prefit SM background predictions (filled solid areas)
  for the 10 aggregate search regions,
  where ``prefit'' means there is no constraint from the likelihood fit.
  The lower panel shows the ratio of the data to
  the total background prediction.
  The hatched bands correspond to the total uncertainty in the
  background prediction.
   }
    \label{fig:aggSearchBinResults}

\end{figure}

The number of observed events
and the predicted number of SM background events
in each of the 84 search regions
are summarized in Fig.~\ref{fig:baseline_SR}.
Numerical values are given in
Tables~\ref{tab:obs_vs_pred_p1}--\ref{tab:obs_vs_pred_p2}
of Appendix~\ref{src:appendixA}.
The corresponding results for the aggregate search regions are
presented in Fig.~\ref{fig:aggSearchBinResults},
with numerical values in
Table~\ref{tab:agg_sb_obs_pred} of Appendix~\ref{src:appendixA}.
No statistically significant deviation between the
data and the background predictions is observed.
The largest source of background typically arises from
\ttbar or {\wjets} production,
followed by {\znunujets} production.
The latter background source can be dominant,
however,
in search regions with a high \ptmiss threshold.
The contributions of the QCD multijet and rare backgrounds are
small in all regions.

Exclusion limits are derived for
the signal models of Section~\ref{sec:models}
using a binned likelihood fit to the data.
The likelihood function is given by the product of Poisson
probability density functions,
one for each search region and for each of the corresponding
regions of the single-electron, single-muon, and
QCD data control samples,
that account for the
background predictions and signal yields.
The uncertainties are treated as nuisance parameters
with log-normal probability density functions.
Correlations between search regions are taken into account.
Upper limits at 95\% confidence level (CL) on the SUSY production cross
sections are calculated using a modified frequentist approach with the
CL$_\text{s}$ criterion~\cite{Junk:1999kv,Read:2002hq}
and asymptotic results for the test statistic~\cite{CLs,Cowan:2010js}.
Signal models for which the 95\% CL upper limit
on the production cross section falls below the theoretical cross section
(based on NLO+NLL calculations~\cite{Borschensky:2014cia})
are considered to be excluded by the analysis.

The uncertainties in the signal modeling are determined
individually for each search region and account for the following sources:
the statistical uncertainty in the simulated event samples,
the integrated luminosity (2.5\%~\cite{CMS-PAS-LUM-17-001}),
the lepton and isolated-track veto efficiencies (up to 6.8\%),
the $\cPqb$ tagging efficiency (up to 21\%),
the trigger efficiency (up to 2.6\%),
the renormalization and factorization scales (up to 3.5\%),
the ISR modeling (up to 46\%),
the jet energy scale corrections (up to 34\%),
the top quark reconstruction efficiency (up to 14\%),
and the modeling of the fast simulation compared with the full simulation
for top quark reconstruction and mistagging (up to 24\%).
All uncertainties
except those from the statistical precision of the simulation
are treated as fully correlated between search regions.
Signal contamination,
namely potential contributions of signal events to the control samples,
is taken into account when computing the limits.
Note that signal contamination is significant only for the single-lepton
control samples of Section~\ref{sec:TopW}
and is negligible for the dimuon and inverted-$\Delta\phi$
control samples of Sections~\ref{sec:Zinv} and~\ref{sec:QCD}.

\begin{figure}[th]
\centering
\includegraphics[width=\cmsFigWidth]{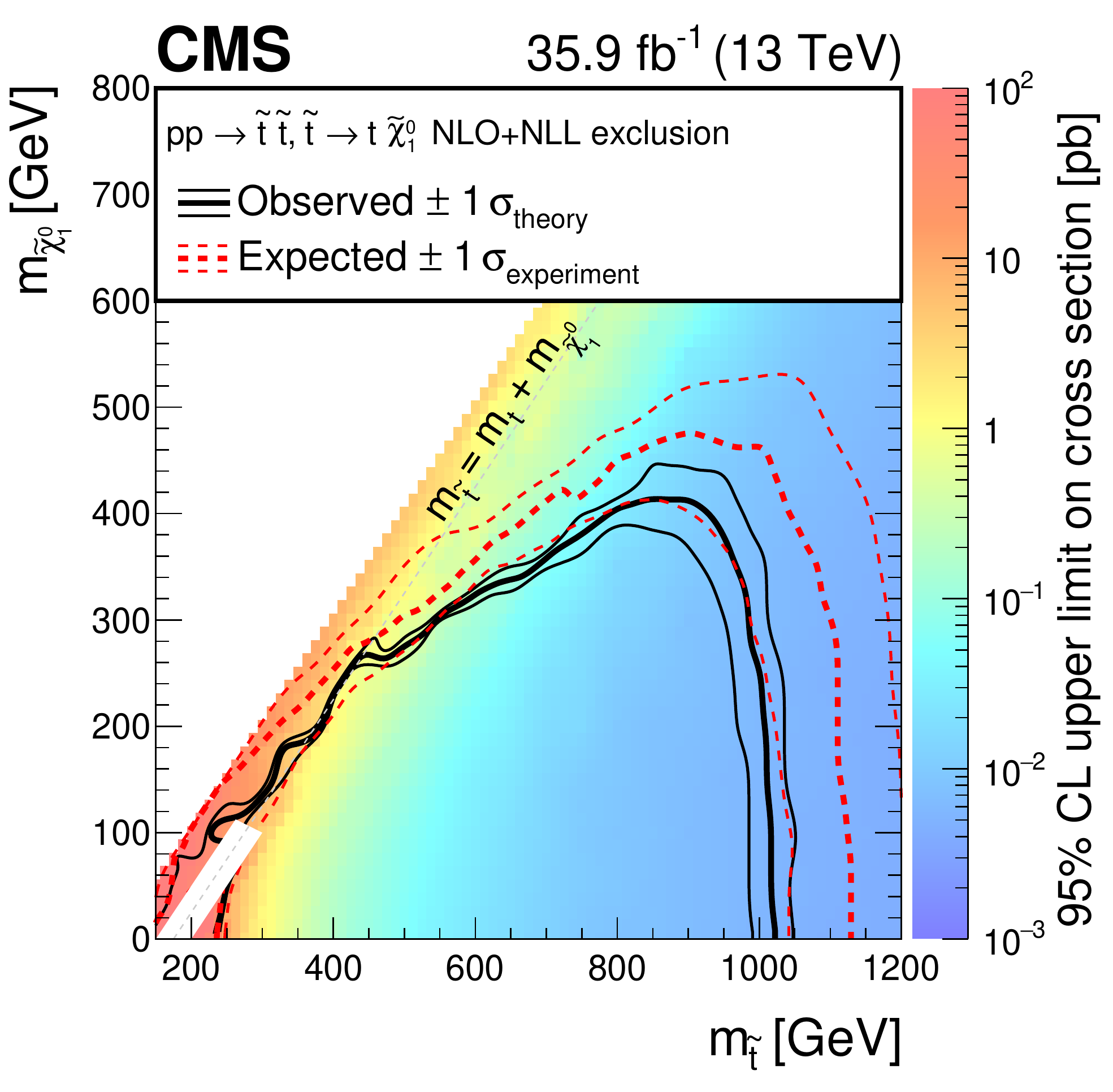}
\caption{
The 95\% CL upper limit on the production cross section
of the T2tt simplified model as a function of the top squark and LSP masses.
The solid black curves represent the observed exclusion contour
with respect to NLO+NLL signal cross sections
and the change in this contour due to variation of these
cross sections within their theoretical uncertainties~\cite{Borschensky:2014cia}.
The dashed red curves indicate the mean expected exclusion contour
and the region containing 68\% of the distribution of expected
exclusion limits under the background-only hypothesis.
No interpretation is provided for signal models for which
$|m_{\stopq} - m_{\lsp} - m_{\cPqt}| \le 25\GeV$ and $m_{\stopq} \leq 275\GeV$
because signal events are essentially indistinguishable from SM
\ttbar events in this region,
rendering the signal event acceptance difficult to model.
}
\label{fig:T2ttlimit}
\end{figure}

\begin{figure*}[!hbt]
\centering
\includegraphics[width=0.48\textwidth]{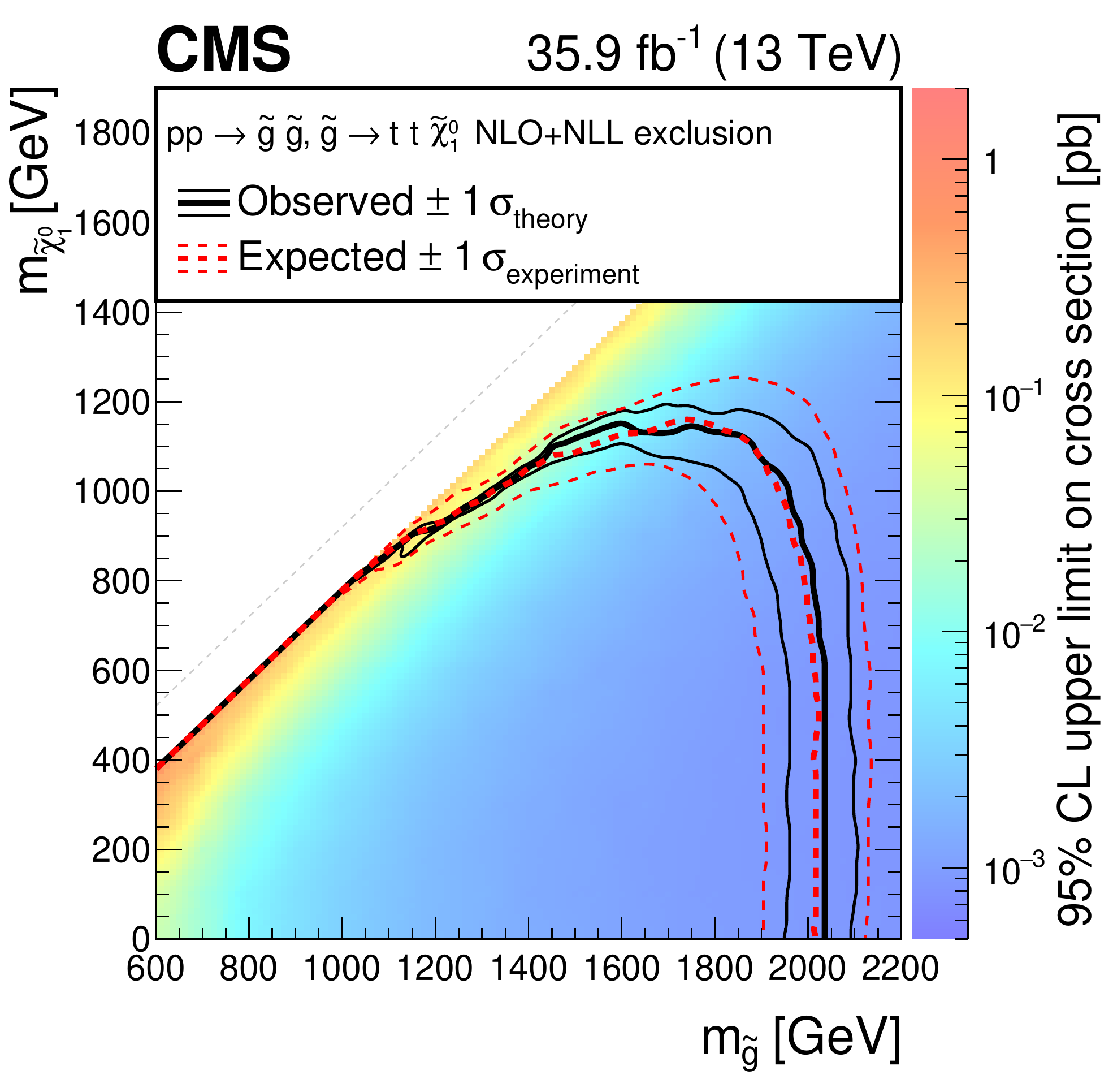}
\includegraphics[width=0.48\textwidth]{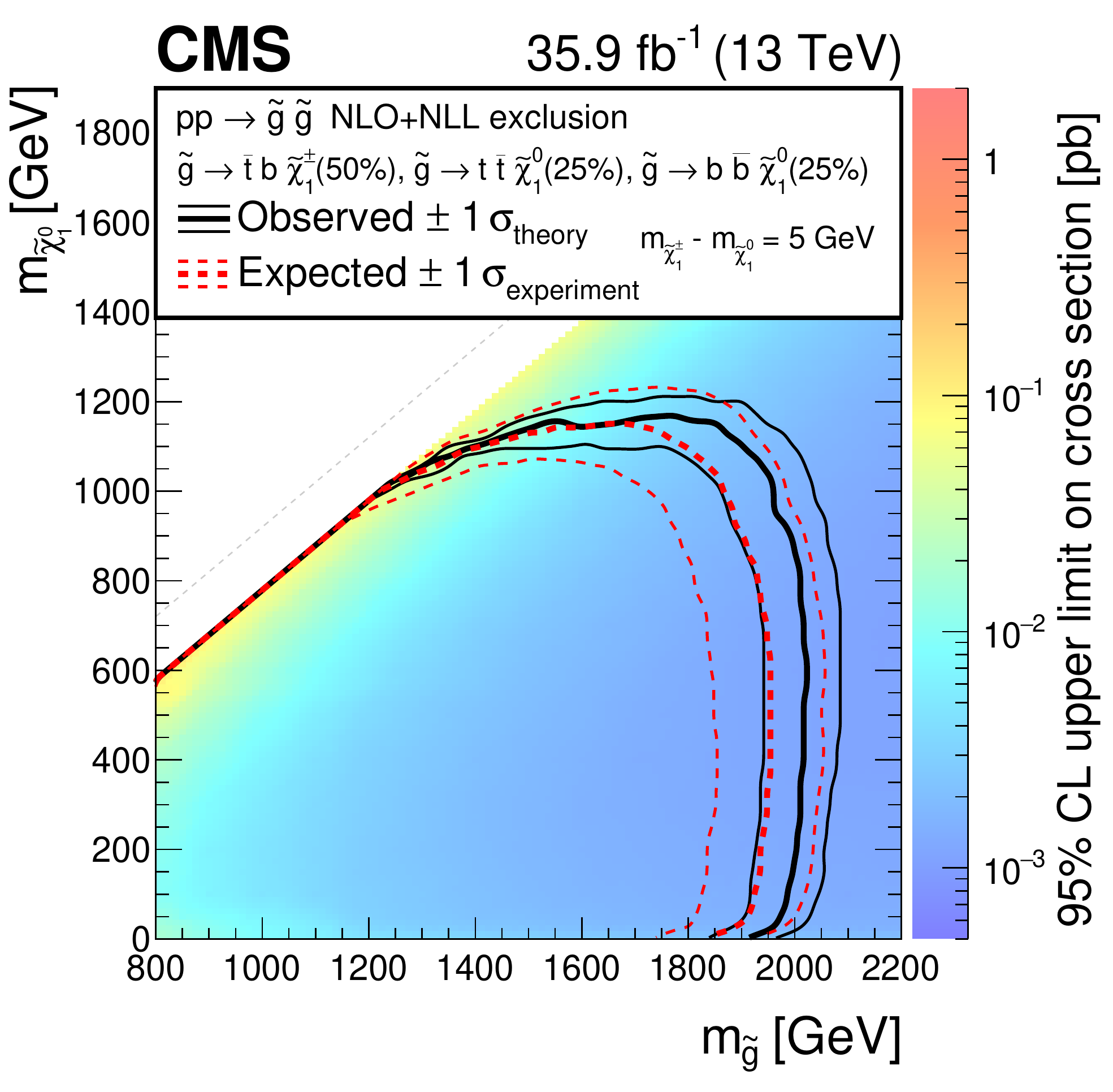}
\includegraphics[width=0.48\textwidth]{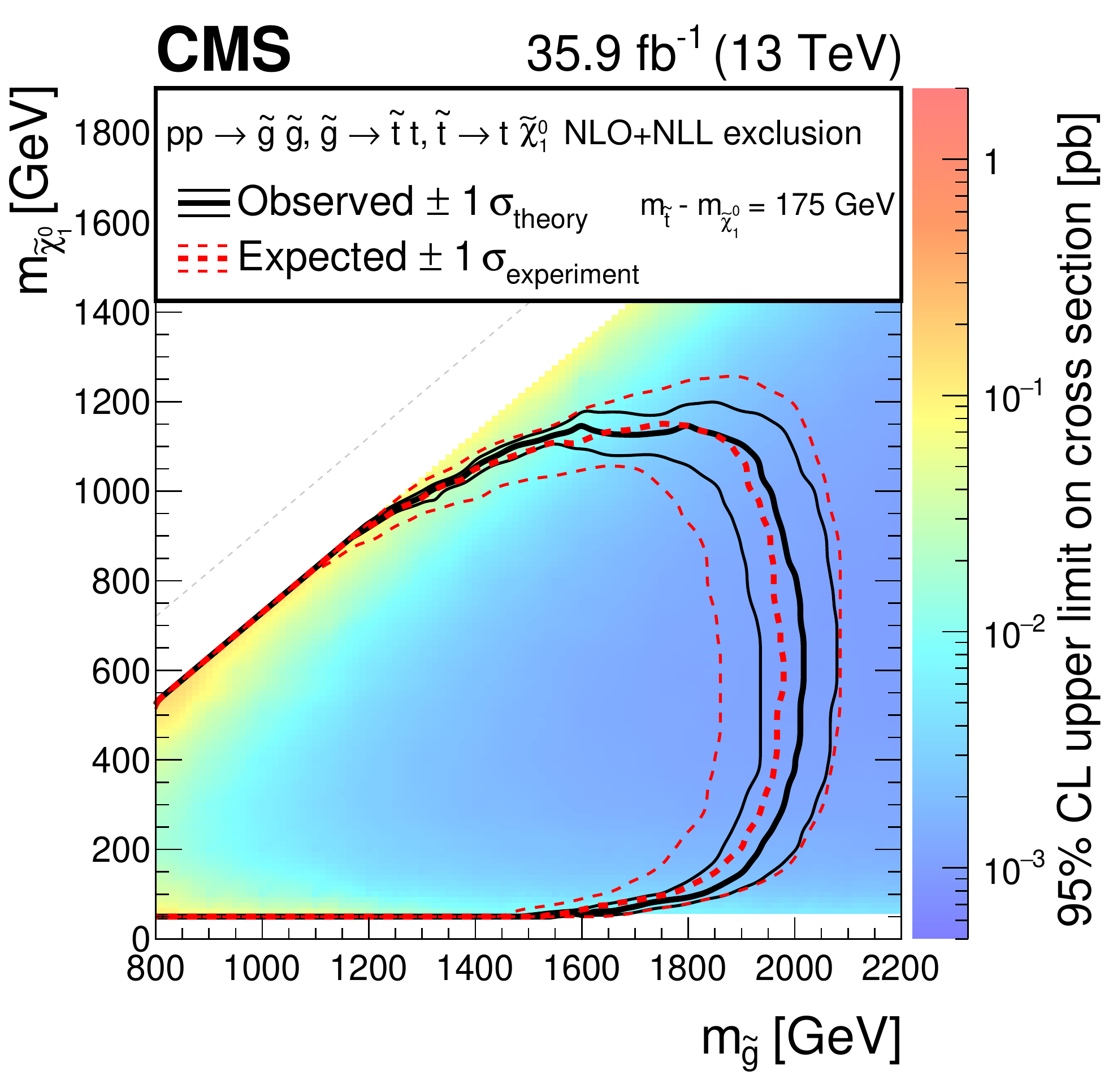}
\includegraphics[width=0.48\textwidth]{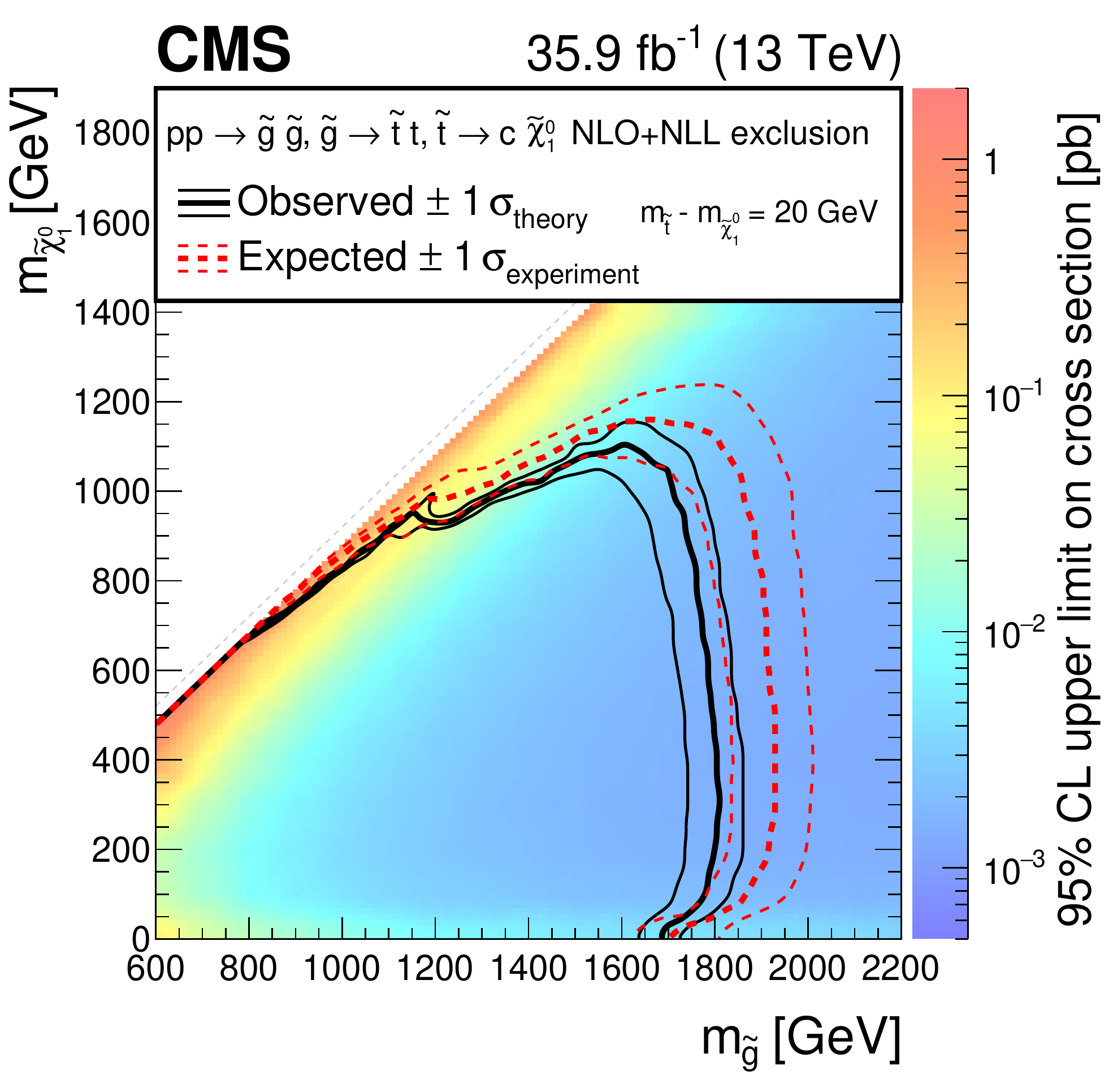}
\caption{
The 95\% CL upper limit on the production cross section of the
T1tttt (upper left),
T1ttbb (upper right),
T5tttt (bottom left), and
T5ttcc (bottom right)
simplified models as a function of the gluino and LSP masses.
The meaning of the curves is explained in the Fig.~\ref{fig:T2ttlimit} caption.
Limits are not given for the T5tttt model for $\mlsp<50\GeV$ for the reason
stated in the text.
}
\label{fig:T5ttXXlimit}
\end{figure*}

Figure~\ref{fig:T2ttlimit} shows the 95\% CL exclusion limits
obtained for the T2tt model of direct top squark pair production:
top squark masses up to 1020\GeV and LSP masses up to 430\GeV are excluded.
The results for the four models of gluino pair production,
T1tttt, T1ttbb, T5tttt, and T5ttcc,
are shown in Fig.~\ref{fig:T5ttXXlimit}.
Gluino masses up to 2040\GeV and LSP masses up to 1150\GeV
are excluded for the T1tttt model,
with corresponding limits of 2020 and 1150\GeV for the T1ttbb model,
2020 and 1150\GeV for the T5tttt model,
and 1810 and 1100\GeV for the T5ttcc model.
The limits on the gluino mass are somewhat lower for the T1ttbb model
than for the T1tttt model
because of the smaller average number of top quarks.
The lower limit of up to 2040\GeV obtained for the gluino mass
in the T1tttt model improves the corresponding
limits of Refs.~\cite{Sirunyan:2017cwe,Sirunyan:2017kqq}
by around 100\GeV,
while the limit on the gluino mass of up to 1810\GeV obtained for the T5ttcc model
improves that presented in Ref.~\cite{Sirunyan:2017uyt}
by 560\GeV.
This emphasizes the effectiveness
of top quark tagging in all-hadronic events as a means
to search for new physics that yields top quarks,
and the complementarity of our study with respect to searches
based on other signatures.

In the case of the T5tttt model there is a significant degradation
of the exclusion limit as \mlsp approaches zero.
This is a consequence of the kinematics of the
$\stopq\to \topq \lsp$ decay,
by which only a small portion of the top squark momentum
is transferred to the LSP if the LSP is lighter than the top quark.
The events then have very small \ptmiss and a small selection efficiency.
The correction to account for signal contamination becomes larger
than the number of selected signal events
and the statistical treatment to account for signal contamination
becomes unreliable.
For this reason,
we do not present results for the T5tttt model if $\mlsp<50\GeV$.

\section{Summary}
\label{sec:summary}

Results are presented from a search for direct and
gluino-mediated top squark production in
proton-proton collisions at a center-of-mass
energy of 13\TeV.
The centerpiece of the analysis is a top quark tagging
algorithm that identifies hadronically decaying top quarks
with high efficiency across a wide range of
top quark transverse momentum~\pt.
The search is based on all-hadronic events with at least four jets,
at least one tagged top quark,
at least one tagged bottom quark jet,
and a large imbalance in transverse momentum \ptmiss.
The data correspond to an integrated luminosity
of 35.9\fbinv
collected with the CMS detector at the LHC in 2016.
A set of 84 search regions is defined based on \ptmiss,
the mass variable \MTTwo,
the scalar \pt sum of jets \HT,
the number of tagged top quarks,
and the number of tagged bottom quark jets.
No statistically significant excess of events is observed relative to
the expectation from the standard model.

Cross section upper limits at 95\% confidence level are evaluated
for a simplified model of direct top squark pair production,
in which the top squarks decay to a top quark
and the lightest supersymmetric particle (LSP) neutralino,
and for simplified models of gluino pair production,
in which the gluinos decay to final states containing
top quarks and LSPs.
Using the signal cross sections calculated with
next-to-leading-order plus next-to-leading-logarithm accuracy,
95\% confidence level lower limits are set on
the masses of the top squark, the gluino, and the LSP.
For the model of direct top squark pair production,
top squark masses up to 1020\GeV
and LSP masses up to 430\GeV are excluded.
For the models of gluino pair production,
gluinos with masses as large as 1810 to 2040\GeV are excluded,
depending on the model,
with corresponding exclusions for
LSPs with masses as large as 1100 to 1150\GeV.
These results significantly extend those
of our previous study~\cite{Khachatryan:2017rhw}.
The use of top quark tagging provides a novel
means to search for new phenomena at the LHC,
yielding complementary sensitivity to other approaches.

\begin{acknowledgments}
\hyphenation{Bundes-ministerium Forschungs-gemeinschaft Forschungs-zentren} We congratulate our colleagues in the CERN accelerator departments for the excellent performance of the LHC and thank the technical and administrative staffs at CERN and at other CMS institutes for their contributions to the success of the CMS effort. In addition, we gratefully acknowledge the computing centers and personnel of the Worldwide LHC Computing Grid for delivering so effectively the computing infrastructure essential to our analyses. Finally, we acknowledge the enduring support for the construction and operation of the LHC and the CMS detector provided by the following funding agencies: the Austrian Federal Ministry of Science, Research and Economy and the Austrian Science Fund; the Belgian Fonds de la Recherche Scientifique, and Fonds voor Wetenschappelijk Onderzoek; the Brazilian Funding Agencies (CNPq, CAPES, FAPERJ, and FAPESP); the Bulgarian Ministry of Education and Science; CERN; the Chinese Academy of Sciences, Ministry of Science and Technology, and National Natural Science Foundation of China; the Colombian Funding Agency (COLCIENCIAS); the Croatian Ministry of Science, Education and Sport, and the Croatian Science Foundation; the Research Promotion Foundation, Cyprus; the Ministry of Education and Research, Estonian Research Council via IUT23-4 and IUT23-6 and European Regional Development Fund, Estonia; the Academy of Finland, Finnish Ministry of Education and Culture, and Helsinki Institute of Physics; the Institut National de Physique Nucl\'eaire et de Physique des Particules~/~CNRS, and Commissariat \`a l'\'Energie Atomique et aux \'Energies Alternatives~/~CEA, France; the Bundesministerium f\"ur Bildung und Forschung, Deutsche Forschungsgemeinschaft, and Helmholtz-Gemeinschaft Deutscher Forschungszentren, Germany; the General Secretariat for Research and Technology, Greece; the National Scientific Research Foundation, and National Innovation Office, Hungary; the Department of Atomic Energy and the Department of Science and Technology, India; the Institute for Studies in Theoretical Physics and Mathematics, Iran; the Science Foundation, Ireland; the Istituto Nazionale di Fisica Nucleare, Italy; the Ministry of Science, ICT and Future Planning, and National Research Foundation (NRF), Republic of Korea; the Lithuanian Academy of Sciences; the Ministry of Education, and University of Malaya (Malaysia); the Mexican Funding Agencies (CINVESTAV, CONACYT, SEP, and UASLP-FAI); the Ministry of Business, Innovation and Employment, New Zealand; the Pakistan Atomic Energy Commission; the Ministry of Science and Higher Education and the National Science Centre, Poland; the Funda\c{c}\~ao para a Ci\^encia e a Tecnologia, Portugal; JINR, Dubna; the Ministry of Education and Science of the Russian Federation, the Federal Agency of Atomic Energy of the Russian Federation, Russian Academy of Sciences, and the Russian Foundation for Basic Research; the Ministry of Education, Science and Technological Development of Serbia; the Secretar\'{\i}a de Estado de Investigaci\'on, Desarrollo e Innovaci\'on and Programa Consolider-Ingenio 2010, Spain; the Swiss Funding Agencies (ETH Board, ETH Zurich, PSI, SNF, UniZH, Canton Zurich, and SER); the Ministry of Science and Technology, Taipei; the Thailand Center of Excellence in Physics, the Institute for the Promotion of Teaching Science and Technology of Thailand, Special Task Force for Activating Research and the National Science and Technology Development Agency of Thailand; the Scientific and Technical Research Council of Turkey, and Turkish Atomic Energy Authority; the National Academy of Sciences of Ukraine, and State Fund for Fundamental Researches, Ukraine; the Science and Technology Facilities Council, UK; the US Department of Energy, and the US National Science Foundation.

Individuals have received support from the Marie-Curie program and the European Research Council and EPLANET (European Union); the Leventis Foundation; the A. P. Sloan Foundation; the Alexander von Humboldt Foundation; the Belgian Federal Science Policy Office; the Fonds pour la Formation \`a la Recherche dans l'Industrie et dans l'Agriculture (FRIA-Belgium); the Agentschap voor Innovatie door Wetenschap en Technologie (IWT-Belgium); the Ministry of Education, Youth and Sports (MEYS) of the Czech Republic; the Council of Science and Industrial Research, India; the HOMING PLUS program of Foundation for Polish Science, cofinanced from European Union, Regional Development Fund; the Compagnia di San Paolo (Torino); the Consorzio per la Fisica (Trieste); MIUR project 20108T4XTM (Italy); the Thalis and Aristeia programs cofinanced by EU-ESF and the Greek NSRF; and the National Priorities Research Program by Qatar National Research Fund.
\end{acknowledgments}

\bibliography{auto_generated}

\clearpage
\appendix

\section{Prefit background predictions}
\label{src:appendixA}

Tables~\ref{tab:obs_vs_pred_p1}--\ref{tab:obs_vs_pred_p2}
present the prefit predictions
for the number of standard model background events
in each of the 84 search regions,
along with the number of observed events.
``Prefit" means that there is no constraint from the likelihood fit.
The corresponding information for the 10 aggregate search regions
is presented in Table~\ref{tab:agg_sb_obs_pred}.

\begin{table*}[ht]
\centering
\topcaption{
The observed number of events and the total background prediction
for search regions with $\ntops=1$ and $\nbjets=1$.
The first uncertainty in the background prediction is
statistical and the second is systematic.
}
\label{tab:obs_vs_pred_p1}
\renewcommand{\arraystretch}{1.15}
\begin{scotch}{ccccc  cc}
     Search region &          \ntops &         \nbjets &   \MTTwo [\GeVns{}] &     \met [\GeVns{}]  & Data   & Predicted background \\
 \hline
  1 &          1 &          1 &    200--300 &    250--400 &       1649 &  $1600\pm 30^{+  130} _{-  140}$ \\
  2 &          1 &          1 &    200--300 &    400--500 &         85 &  $73^{+    7}_{-    6}$ $^{+   12} _{-    9}$ \\
  3 &          1 &          1 &    200--300 &    500--600 &         23 &  $18^{+    4} _{-    3}$ $^{+    6} _{-    4}$ \\
  4 &          1 &          1 &    200--300 &    600--750 &          7 &  $3.6^{+  1.9} _{-  0.8}$ $^{+  1.9} _{-  0.8}$ \\
  5 &          1 &          1 &    200--550 &   $\geq$750 &        7 &  $5.0^{+  2.4} _{-  1.1}$ $^{+  1.9} _{-  1.2}$ \\
  6 &          1 &          1 &    300--400 &    250--400 &       1020 &  $890\pm 20^{+   80} _{-   80}$ \\
  7 &          1 &          1 &    300--400 &    400--500 &         87 &  $79^{+    7} _{-    6}\pm 9$ \\
  8 &          1 &          1 &    300--400 &    500--600 &         23 &  $17^{+    4} _{-    2}\pm 3$ \\
  9 &          1 &          1 &    300--400 &    600--750 &          9 &  $3.7^{+  2.2} _{-  0.8}$ $^{+  1.6} _{-  0.9}$ \\
 10 &          1 &          1 &    400--550 &    250--400 &        108 &  $107^{+    8} _{-    7}\pm 10$ \\
 11 &          1 &          1 &    400--550 &    400--500 &        116 &  $105^{+    7} _{-    6}\pm 10$ \\
 12 &          1 &          1 &    400--550 &    500--600 &         47 &  $38^{+    5} _{-    4}\pm 7$ \\
 13 &          1 &          1 &    400--550 &    600--750 &         12 &  $8.1^{+  2.4} _{-  1.2}\pm 1.9$ \\
 14 &          1 &          1 &    550--750 &    250--400 &          1 &  $0.7^{+  1.0} _{-  0.3}$ $^{+  0.4} _{-  0.2}$ \\
 15 &          1 &          1 &    550--750 &    400--500 &          7 &  $4.3^{+  2.0} _{-  1.1}\pm 0.8$ \\
 16 &          1 &          1 &    550--750 &    500--600 &         17 &  $13^{+    3} _{-    2}\pm 3$ \\
 17 &          1 &          1 &    550--750 &    600--750 &         10 &  $19^{+    3} _{-    2}\pm 4$ \\
 18 &          1 &          1 &    550--750 &   $\geq$750 &        7 &  $4.0^{+  1.5} _{-  0.3}\pm 1.8$ \\
 19 &          1 &          1 &   $\geq$750 &  250--600 &          0 &  $0.1^{+ 1.7} _{- 0.1}\pm 0.1$ \\
 20 &          1 &          1 &   $\geq$750 &  600--750 &          1 &  $1.9^{+  2.2} _{-  1.0}$ $^{+  0.9} _{-  0.8}$ \\
 21 &          1 &          1 &   $\geq$750 & $\geq$750 &        8 &  $4.6^{+  1.6} _{-  0.5}\pm 1.9$ \\
\end{scotch}
\end{table*}

\begin{table*}[tbh]
\centering
\topcaption{
The observed number of events and the total background prediction
for search regions with $\ntops=1$ and $\nbjets\geq2$.
The first uncertainty in the background prediction is
statistical and the second is systematic.
}
\label{tab:obs_vs_pred_p11}
\renewcommand{\arraystretch}{1.15}
\begin{scotch}{ccccc  cc}
     Search region &          \ntops &         \nbjets &   \MTTwo [\GeVns{}] &     \met [\GeVns{}]  & Data   & Predicted background \\
 \hline
 22 &          1 &          2 &    200--350 &    250--400 &        596 &   580 $\pm 20\pm 60$ \\
 23 &          1 &          2 &    200--350 &    400--500 &         59 &    41 $^{+    6} _{-    5}$ $^{+    6} _{-    5}$ \\
 24 &          1 &          2 &    200--350 &    500--600 &         14 &   8.7 $^{+  3.4} _{-  2.1}\pm 1.3$ \\
 25 &          1 &          2 &    200--350 &    600--750 &          2 &   2.1 $^{+  2.7} _{-  0.8}\pm 0.5$ \\
 26 &          1 &          2 &    200--650 &      $\geq$750 &          1 &   3.0 $^{+  2.4} _{-  1.0}$ $^{+  0.9} _{-  0.6}$ \\
 27 &          1 &          2 &    350--450 &    250--400 &         69 &    67 $^{+    6} _{-    5}$ $^{+   18} _{-   14}$ \\
 28 &          1 &          2 &    350--450 &    400--500 &         19 &    13 $^{+    4} _{-    2}\pm 3$ \\
 29 &          1 &          2 &    350--450 &    500--600 &          4 &   3.2 $^{+  2.1} _{-  0.9}\pm 1.0$ \\
 30 &          1 &          2 &    350--450 &    600--750 &          2 &   0.6 $^{+  1.4} _{-  0.1}\pm 0.3$ \\
 31 &          1 &          2 &    450--650 &    250--400 &          3 &   4.0 $^{+  2.0} _{-  1.1}$ $^{+  0.7} _{-  0.9}$ \\
 32 &          1 &          2 &    450--650 &    400--500 &          9 &   9.7 $^{+  2.7} _{-  1.8}$ $^{+  2.1} _{-  2.0}$ \\
 33 &          1 &          2 &    450--650 &    500--600 &          6 &   6.0 $^{+  1.6} _{-  0.9}\pm 1.9$ \\
 34 &          1 &          2 &    450--650 &    600--750 &          2 &   4.6 $^{+  2.6} _{-  1.3}\pm 1.2$ \\
 35 &          1 &          2 &   $\geq$650 &    250--600 &        0 &  0.06 $^{+ 1.03} _{- 0.03}\pm 0.03$ \\
 36 &          1 &          2 &   $\geq$650 &    600--750 &        0 &   1.0 $^{+  1.8} _{-  0.1}\pm 0.5$ \\
 37 &          1 &          2 &   $\geq$650 &  $\geq$750 &       2 &   1.2 $^{+  1.1} _{-  0.1}\pm 0.5$ \\
 38 &          1 &        $\geq$3 &   300--1000 &    250--350 &   85 &    81 $^{+    9} _{-    8}\pm 7$ \\
 39 &          1 &        $\geq$3 &   300--1000 &    350--450 &   22 &    15 $^{+    5} _{-    3}\pm 2$ \\
 40 &          1 &        $\geq$3 &   300--1000 &    450--550 &    6 &   4.5 $^{+  3.4} _{-  1.7}\pm 0.8$ \\
 41 &          1 &        $\geq$3 &   300--1000 & $\geq$550 &    2 &   2.4 $^{+  2.9} _{-  1.0}$ $^{+  1.0} _{-  0.7}$ \\
 42 &          1 &        $\geq$3 &  1000--1500 &    250--350 &   12 &    13 $^{+    4} _{-    3}\pm 2$ \\
 43 &          1 &        $\geq$3 &  1000--1500 &    350--450 &    5 &   5.0 $^{+  2.7} _{-  1.7}\pm 1.1$ \\
 44 &          1 &        $\geq$3 &  1000--1500 &    450--550 &    0 &   1.8 $^{+  2.3} _{-  1.0}\pm 0.4$ \\
 45 &          1 &        $\geq$3 &  1000--1500 & $\geq$550 &    3 &   2.7 $^{+  3.9} _{-  1.4}$ $^{+  0.6} _{-  0.5}$ \\
 46 &          1 &        $\geq$3 & $\geq$1500 &    250--350 &   2 &   9.6 $^{+  3.4} _{-  2.2}\pm 3.3$ \\
 47 &          1 &        $\geq$3 & $\geq$1500 &    350--550 &   1 &   3.4 $^{+  2.3} _{-  1.2}$ $^{+  3.4} _{-  1.5}$ \\
 48 &          1 &        $\geq$3 & $\geq$1500 & $\geq$550 &   0 &   1.3 $^{+  1.8} _{-  0.7}\pm 0.3$ \\
\end{scotch}
\end{table*}

\begin{table*}[tbp]
\centering
\topcaption{
The observed number of events and the total background prediction
for search regions with $\ntops\geq2$.
The first uncertainty in the background prediction is
statistical and the second is systematic.
}
\label{tab:obs_vs_pred_p2}
\renewcommand{\arraystretch}{1.15}
\begin{scotch}{ccccc  cc}
     Search region &          \ntops &         \nbjets &   \MTTwo [\GeVns{}] &     \met [\GeVns{}]  &  Data   & Predicted background \\
 \hline
 49 &          2 &          1 &    200--300 &    250--350 &         57 &    60 $^{+    6} _{-    5}\pm 11$ \\
 50 &          2 &          1 &    200--300 &    350--450 &         10 &   7.5 $^{+  2.5} _{-  1.7}$ $^{+  1.8} _{-  1.4}$ \\
 51 &          2 &          1 &    200--300 &    450--600 &          0 &   2.2 $^{+  1.4} _{-  0.8}$ $^{+  0.8} _{-  0.5}$ \\
 52 &          2 &          1 &    200--450 & $\geq$600 &          0 &   0.9 $^{+  2.0} _{-  0.6}$ $^{+  0.4} _{-  0.3}$ \\
 53 &          2 &          1 &    300--450 &    250--350 &         38 &    32 $^{+    5} _{-    4}\pm 3$ \\
 54 &          2 &          1 &    300--450 &    350--450 &          8 &    11 $^{+    3} _{-    2}\pm 2$ \\
 55 &          2 &          1 &    300--450 &    450--600 &          4 &   2.1 $^{+  1.7} _{-  0.7}$ $^{+  0.8} _{-  0.5}$ \\
 56 &          2 &          1 & $\geq$450 &    250--450 &          2 &   1.8 $^{+  1.5} _{-  0.6}\pm 0.4$ \\
 57 &          2 &          1 & $\geq$450 &    450--600 &          3 &   3.3 $^{+  2.7} _{-  1.1}\pm 0.9$ \\
 58 &          2 &          1 & $\geq$450 & $\geq$600 &          7 &   1.0 $^{+  1.2} _{-  0.1}\pm 0.5$ \\
 59 &          2 &          2 &    200--300 &    250--350 &         46 &    43 $\pm 5^{+    5} _{-    6}$ \\
 60 &          2 &          2 &    200--300 &    350--450 &         11 &   8.7 $^{+  2.7} _{-  1.9}$ $^{+  1.4} _{-  1.3}$ \\
 61 &          2 &          2 &    200--300 &    450--600 &          1 &   0.6 $^{+  1.6} _{-  0.4}$ $^{+  0.3} _{-  0.2}$ \\
 62 &          2 &          2 &    200--400 & $\geq$600 &          1 &   0.6 $^{+  1.7} _{-  0.5}\pm 0.2$ \\
 63 &          2 &          2 &    300--400 &    250--350 &         28 &    27 $^{+    5} _{-    4}\pm 3$ \\
 64 &          2 &          2 &    300--400 &    350--450 &          6 &   4.9 $^{+  2.9} _{-  1.6}\pm 0.9$ \\
 65 &          2 &          2 &    300--400 &    450--600 &          3 &   1.7 $^{+  2.4} _{-  1.0}$ $^{+  0.6} _{-  0.5}$ \\
 66 &          2 &          2 &    400--500 &    250--450 &          4 &   4.7 $^{+  2.3} _{-  1.2}$ $^{+  0.7} _{-  0.8}$ \\
 67 &          2 &          2 &    400--500 &    450--600 &          1 &   1.4 $^{+  2.7} _{-  0.7}$ $^{+  0.4} _{-  0.6}$ \\
 68 &          2 &          2 & $\geq$400 & $\geq$600 &          1 &  0.5 $^{+ 2.7} _{- 0.1}\pm 0.2$ \\
 69 &          2 &          2 & $\geq$500 &    250--450 &          0 &  0.1 $^{+ 1.4} _{- 0.1}\pm 0.1$ \\
 70 &          2 &          2 & $\geq$500 &    450--600 &          2 &  0.5 $^{+ 2.2} _{- 0.1}\pm 0.1$ \\
 71 &          2 &  $\geq$3 &    300--900 &    250--350 &          3 &   9.6 $^{+  3.0} _{-  2.1}\pm 1.7$ \\
 72 &          2 &  $\geq$3 &    300--900 &    350--500 &          2 &   0.7 $^{+  2.0} _{-  0.4}\pm 0.2$ \\
 73 &          2 &  $\geq$3 &   300--1300 & $\geq$500 &          0 &   0.3 $^{+  0.5} _{-  0.3}$ $^{+  0.3} _{-  0.2}$ \\
 74 &          2 &  $\geq$3 &   900--1300 &    250--350 &          6 &   4.7 $^{+  2.9} _{-  1.7}$ $^{+  0.7} _{-  0.9}$ \\
 75 &          2 &  $\geq$3 &   900--1300 &    350--500 &          3 &   1.2 $^{+  1.6} _{-  0.7}\pm 0.4$ \\
 76 &          2 &  $\geq$3 &$\geq$1300 &    250--350 &          3 &   3.5 $^{+  2.1} _{-  1.2}\pm 1.4$ \\
 77 &          2 &  $\geq$3 &$\geq$1300 &    350--500 &          2 &   2.1 $^{+  2.1} _{-  1.0}$ $^{+  0.4} _{-  0.5}$ \\
 78 &          2 &  $\geq$3 &$\geq$1300 & $\geq$500 &          0 &   0.2 $^{+  1.7} _{-  0.3}\pm 0.2$ \\
 79 &  $\geq$3 &        1 &   $\geq$300 &    250--350 &          0 &   0.3 $^{+  2.0} _{-  0.3}\pm 0.2$ \\
 80 &  $\geq$3 &        1 &   $\geq$300 & $\geq$350 &          1 &   0.6 $^{+  1.6} _{-  0.5}\pm 0.2$ \\
 81 &  $\geq$3 &        2 &   $\geq$300 &    250--400 &          1 &   1.7 $^{+  1.5} _{-  0.7}$ $^{+  0.6} _{-  0.5}$ \\
 82 &  $\geq$3 &        2 &   $\geq$300 & $\geq$400 &          0 &  0.1 $^{+ 2.2} _{- 0.1}\pm 0.1$ \\
 83 &  $\geq$3 &$\geq$3 &   $\geq$300 &    250--350 &          0 &  0.5 $^{+  1.5} _{-  0.4}\pm 0.5$ \\
 84 &  $\geq$3 &$\geq$3 &   $\geq$300 & $\geq$350 &          0 &  0.0 $^{+ 1.6} _{-  0.0}$ $^{+  0.1} _{-  0.0}$ \\
\end{scotch}
\end{table*}

\begin{table*}[htb]
\centering
\topcaption{
The observed number of events and the total background prediction
for the aggregate search regions.
The first uncertainty in the background prediction is
statistical and the second is systematic.
}
\label{tab:agg_sb_obs_pred}
\renewcommand{\arraystretch}{1.30}
\begin{scotch}{cccccccccccc}
     Search region &  \ntops &    \nbjets &   \MTTwo [\GeVns{}] &     \met [\GeVns{}]  & Data & Predicted background \\
\hline
    1 & $\geq$1 &  $\geq$1 &  $\geq$200 &  $\geq$250  &  4424 &   $4100\pm 50^{ +390}_{ -340}$ \\
    2 & $\geq$2 &  $\geq$2 &  $\geq$200 &  $\geq$250  &   124 &   $116\pm 8^{  +15} _{  -12}$ \\
    3 & $\geq$3 &  $\geq$1 &  $\geq$200 &  $\geq$250  &     2 &   $3.3^{ +2.0} _{ -1.1}$ $^{ +1.2} _{ -1.1}$ \\
    4 & $\geq$3 &  $\geq$3 &  $\geq$200 &  $\geq$250  &     0 &   $0.5^{ +1.4} _{ -0.4}\pm 0.5$ \\
    5 & $\geq$2 &  $\geq$1 &  $\geq$200 &  $\geq$400  &    41 &   $30^{   +4} _{   -3}$ $^{   +5} _{   -4}$ \\
    6 & $\geq$1 &  $\geq$2 &  $\geq$600 &  $\geq$400  &     4 &   $7.5^{ +2.1} _{ -1.2}$ $^{ +2.0} _{ -1.9}$ \\
 \hline
     Search region &  \ntops &    \nbjets &   \HT [\GeVns{}] &     \met [\GeVns{}]  &  Data & Predicted background \\
\hline
    7 & $\geq$1 &  $\geq$2 &  $\geq$1400 & $\geq$500  &     6 &   $6.0^{ +2.7} _{ -1.5}\pm 1.5$ \\
    8 & $\geq$2 &  $\geq$3 &  $\geq$600  & $\geq$350  &     7 &   $3.9^{ +2.1} _{ -1.2}\pm 0.9$ \\
    9 & $\geq$2 &  $\geq$3 &  $\geq$300  & $\geq$500  &     0 &   $0.6^{ +1.0} _{ -0.4}\pm 0.4$ \\
   10 & $\geq$2 &  $\geq$3 &  $\geq$1300 & $\geq$500  &     0 &   $0.2^{ +1.8} _{ -0.3}\pm 0.2$ \\
     \end{scotch}
\end{table*}

\cleardoublepage \section{The CMS Collaboration \label{app:collab}}\begin{sloppypar}\hyphenpenalty=5000\widowpenalty=500\clubpenalty=5000\textbf{Yerevan Physics Institute,  Yerevan,  Armenia}\\*[0pt]
A.M.~Sirunyan, A.~Tumasyan
\vskip\cmsinstskip
\textbf{Institut f\"{u}r Hochenergiephysik,  Wien,  Austria}\\*[0pt]
W.~Adam, F.~Ambrogi, E.~Asilar, T.~Bergauer, J.~Brandstetter, E.~Brondolin, M.~Dragicevic, J.~Er\"{o}, A.~Escalante Del Valle, M.~Flechl, M.~Friedl, R.~Fr\"{u}hwirth\cmsAuthorMark{1}, V.M.~Ghete, J.~Grossmann, J.~Hrubec, M.~Jeitler\cmsAuthorMark{1}, A.~K\"{o}nig, N.~Krammer, I.~Kr\"{a}tschmer, D.~Liko, T.~Madlener, I.~Mikulec, E.~Pree, N.~Rad, H.~Rohringer, J.~Schieck\cmsAuthorMark{1}, R.~Sch\"{o}fbeck, M.~Spanring, D.~Spitzbart, W.~Waltenberger, J.~Wittmann, C.-E.~Wulz\cmsAuthorMark{1}, M.~Zarucki
% ...
\vskip\cmsinstskip
% ...
1:~~Also at Vienna University of Technology, Vienna, Austria\\
% ...

\end{sloppypar}
\end{document}